\documentclass[11pt]{article}

\usepackage[hyphens]{url}  
\usepackage{hyperref}
\usepackage{breakurl} 

\usepackage[toc,page,header]{appendix}
\usepackage{titletoc}

\usepackage{geometry}
\usepackage{xcolor}
\usepackage{xspace}
\usepackage[T1]{fontenc}
\usepackage[utf8x]{inputenc}
\usepackage{microtype}

\usepackage[backend=biber,style=numeric-comp,sorting=none,giveninits=true]{biblatex}

\usepackage{authblk}
\usepackage{enumitem}

\renewbibmacro{in:}{}

\title{\vspace*{3cm}
  {\bf Polish national input to the 2026 update}\\
  {\bf of the European Strategy for Particle Physics}}  

\date{March, 2025}

\author[1]{K.~Adamczyk}     
\author[2]{B.~Badełek}     
\author[3]{T.~Banaszkiewicz}     
\author[4]{V.~Batozskaya}     
\author[4]{M.~Bluj}     
\author[1]{D.~Bocian}     
\author[1]{P.~Bruckman de Renstrom}     
\author[3]{M.~Chorowski}     
\author[1]{M.~Chrząszcz}     
\author[5]{W.~Cichalewski}     
\author[1]{D.~Derendarz}     
\author[6]{M.~Dyndał}     
\author[*7]{J.~Gluza}     
\author[8]{Ł.~Graczykowski}     
\author[8]{K.~Grebieszkow}     
\author[*1]{A.~Kaczmarska}     
\author[6,9]{P.~Kalaczyński}     
\author[2]{A.~Kalinowski}     
\author[2]{J.~Kalinowski}     
\author[2]{M.~Konecki}     
\author[*8]{G.~Kornakov}     
\author[6]{P.~Kotko}     
\author[1]{M.~Kucharczyk }     
\author[4]{A.~Kupść}     
\author[1]{A.~Kusina}     
\author[1]{K.~Kutak}     
\author[*4]{J.~Łagoda}     
\author[*1]{T.~Lesiak}     
\author[1]{P.~Malecki}     
\author[1]{A.~Matyja}     
\author[6]{B.~Mindur}     
\author[2]{M.~Misiak}     
\author[4]{A.~Padee}     
\author[2]{L.W.~Piotrowski}     
\author[6]{K.~Piotrzkowski}     
\author[*10]{W.~Płaczek}     
\author[8]{K.~Poźniak}     
\author[2]{K.~Rolbiecki}     
\author[1]{A.~Rybicki}     
\author[1]{S.~Sapeta}     
\author[10]{A.~Siódmok}     
\author[1]{M.~Skrzypek}     
\author[1]{M.~Sławińska}     
\author[11]{J.~Sobczyk}     
\author[1]{I.~Sputowska}     
\author[1]{E.~Stanecka}     
\author[10]{Ł.~Stawarz}     
\author[4]{A.~Szabelski}     
\author[4]{M.~Szleper}     
\author[4]{P.~Sznajder}     
\author[*6]{T.~Szumlak}     
\author[6]{A.~Ukleja}     
\author[1]{T.~Wąchała}     
\author[4]{J.~Wagner}     
\author[*4]{W.~Wiślicki}     
\author[4]{S.~Wronka}     
\author[1]{B.~Żabiński }     
\author[*2]{A.F.~Żarnecki}     
\author[1]{G.~Żarnecki}     
\author[8]{H.~Zbroszczyk}     
\author[6]{A.~Zemła}     
\author[10]{G.~Zuzel}     

\affil[*]{Coordination group}
\affil[1]{The Henryk Niewodniczański Institute of Nuclear Physics Polish Academy of Sciences}
\affil[2]{Faculty of Physics, University of Warsaw}
\affil[3]{Wrocław University of Technology}
\affil[4]{National Centre for Nuclear Research}
\affil[5]{Łódź University of Technology}
\affil[6]{AGH University of Krakow}
\affil[7]{University of Silesia}
\affil[8]{Warsaw University of Technology}
\affil[9]{Nicolaus Copernicus Astronomical Center  Physics Polish Academy of Sciences}
\affil[10]{Faculty of Physics, Astronomy and Applied Computer Science, Jagiellonian University}
\affil[11]{University of Wroclaw}

\newcommand{\ee}{e$^+$e$^-$\xspace}
\newcommand{\as}{\alpha_{\rm s}}

\begin{document}

\maketitle
\thispagestyle{empty}

\vspace*{3cm}

\setcounter{page}{0}

\renewcommand{\thesection}{\Roman{section}}
\renewcommand{\thepage}{\roman{page}}

\newpage


  \section*{Executive summary}

The Polish high energy physics (HEP) community fully recognizes the urgent need to host at CERN a~flagship project implementing a broad, long-term, and comprehensive vision of particle physics research and pursuing technological advances.
Thus, we give preference and declare willingness to actively engage and participate in every aspect of the FCC project (both FCC-ee and FCC-hh), particularly accelerator development, detector construction, theoretical calculations, and physics analyses.
As the \ee Higgs Factory is the top priority for our field, the proposal to build a linear collider facility at CERN, opening up complementary physics prospects, should be considered as the second option.

Polish teams declare strong support and are fully committed to contribute to the full exploitation of all aspects of the physics potential of the LHC and the HL-LHC programmes.
%
To ensure the long-term development of particle physics, we also support the continuation of the high-field magnet research programme, as well as investigating other scenarios including, in particular, linear acceleration techniques and new acceleration technologies such as plasma acceleration, the muon collider and Gamma Factory.
In addition, CERN should continue to provide support to fixed-target programmes at SPS as well as other non-collider and non-accelerator experiments at CERN. Participation in major projects conducted in and outside Europe should also be fostered.

Education, communication, and outreach of particle physics are of paramount importance for the future of our field. An increased effort coordinated at the European level and resources allocated in all Member States are essential to effectively support future large-scale particle physics projects.

\newpage


\section*{I~~~Scope of the document}

Included in this document are conclusions and recommendations prepared as Polish national input to the Update of the European Strategy for Particle Physics, initiated by the CERN Council in March 2024, as well as the general overview of the Particle Physics research in Poland. 
The work was started by the Polish member of RECFA together with the Polish delegation to the CERN Council as well as present and former ECFA members. 
The document is based on inputs prepared by topical expert teams and collected by the coordinating group. For details of the process and the composition of the groups you are referred to the backup document. 

Particle Physics (PP) research, including both experimental and theoretical activities, is conducted in 18 research institutions -- leading universities and research institutes -- distributed in 8 centres in Poland. 
The number of Polish PP researchers counted in FTE is about 365 staff and 190 PhD and MSc Students. 
This is complemented by about 130 engineers and technicians directly involved in research.
For the full list of academic units and research institutes involved in HEP activities in Poland, along with their acronyms used hereafter, and more details on the community research activities, please refer to the tables included in the backup document.

Presented in the subsequent sections of this document are summaries of activities, achievements and plans of the Polish groups in different topical areas, followed by general conclusions and recommendations.

\section*{II~~~Electroweak and Higgs boson physics}

Polish groups investigate the Higgs boson properties and electroweak physics at the LHC as part of the ATLAS and CMS collaborations. 
They studied the $H$ boson decay to the pair of tau leptons and searched for the Higgs boson pair production signal, important for the determination of the $H$ self-coupling constant, thus for the shape of the Higgs potential. 
These analyses involved statistical combinations of di-Higgs searches and measuring Higgs trilinear self-coupling in both Run 1 and early Run 2. Other studies of Polish groups encompass: Higgs boson couplings to polarised $W$ and $Z$ bosons in the vector boson fusion production and $H\to WW^{*}$ decays, and the CP state of the  Higgs boson in its decays to $\tau$ leptons.
The groups are committed to continue analyses towards observing Higgs boson pair production and Higgs boson properties at the  High-Luminosity LHC (HL-LHC).
Vector bosons studies included also a series of precise measurements of $Z$ boson properties in the ATLAS experiment (like the transverse $Z$-boson momentum distribution, differential distributions of the $Z\to ll \gamma$ and $Z\to ll \gamma\gamma$ processes, the measurement of the effective leptonic weak mixing angle using lepton pairs from $Z$-boson decays). For the electroweak measurements, vector boson scattering was analysed with the same-sign~$WW$ scattering and the $W\to\mu/e \nu$ (leptonic) decays of pair of $W$ bosons. 

One of the key problems of particle physics is a profound understanding of the electroweak symmetry breaking. It can be achieved by probing properties of the Higgs boson and precise determination of electroweak pseudo-observables of the Standard Model (like $\sin^2{\theta_W}$, $m_Z$, $m_W$, $\Gamma_Z$, $\alpha_{QED}$, $\alpha_s$) to be used in global electroweak fits. The only compelling option to perform such measurements with precision surpassing LEP and (HL-)LHC seems to be an $e^+e^-$ collider with high luminosity and operation modes at the $Z$-peak, $ZH$ at 240~GeV, and $WW$ and $t\bar{t}$ thresholds at 180 and 365~GeV, respectively. Such an $e^+e^-$ Higgs, electroweak and flavour factory was already recommended by the 2021 update of the European Strategy for Particle Physics (ESPPU) with the highest priority given to complementing and deepening the successful Higgs boson measurements performed at the LHC and HL-LHC.

Among all the proposed future $e^+e^-$ collider solutions, the Future Circular Collider (FCC-ee) with four to five orders of magnitude higher statistics than LEP is the most ambitious, broad-reaching and demanding project. Collecting nearly $10^{13}$ $e^+e^-$ collisions at $Z$-boson resonance (thus also called the TeraZ $e^+e^-$ mode), the proposed CERN circular collider is expected to improve the precision of determination of electroweak pseudo-observables by one to two orders of magnitude.  
For all other measurements, the FCC-ee TeraZ run offers ultimate sensitivity for versatile programme of electroweak, QCD, and flavour physics, as well as numerous searches for rare processes. 
To meet the aforementioned precision goals, Polish theory groups work on calculations of higher-order radiative corrections and development of Monte Carlo (MC) $e^+e^-$ generators.  Present and future calculations must include mixed QCD and electroweak two- and three-loop corrections for $Z$-boson decay observables, pair production processes ($e^+e^- \to f\bar{f}$, $e^+e^- \to W^+W^-$, $e^+e^- \to ZH$), as well as Krakow MC tools ({\sf KKMCee}, {\sf BHLUMI}, {\sf BHWIDE}, {\sf KoralW/YFSWW/YFSZZ}) for the simulation of multiple photon radiation beyond leading-logarithmic approximation. In addition, the Polish community actively develops central FCC software, methods and tools for theoretical calculations and provides support for the generators mentioned above.


\section*{III~~~Beyond the Standard Model and Dark Matter searches}

Krakow ATLAS groups are actively involved in searches for beyond the Standard Model (BSM) phenomena in the Higgs sector. These include direct searches for additional neutral or charged scalars in the wide range of masses as well as probing for anomalous pair-production of the 125~GeV Higgs. The search for non-resonant Higgs pair production provides constraints on anomalous trilinear Higgs coupling modifier $\kappa_\lambda$, while searching for resonant production directly probes for heavy spin-0 or spin-2 BSM states. Additionally, BSM effects are studied in the Higgs boson couplings to polarised $W$ and $Z$ bosons. All analysis activities are being continued to cover data accumulated in LHC Run 3 as well as extended to the HL-LHC period.

The LHCb Krakow group study the BSM physics in the sector of $b$-quarks, concerning the $Z'$ and $b'$ searches in $Z(\to \mu^+\mu^-) b \bar{b}$ channel as well as BSM decays with displaced vertex signature, such as exotic particles predicted by the Hidden Valley models.
Last but not least, the IFJ PAN group participated in the state-of-the-art right-handed-neutrino global fit within the Gambit Collaboration.

The Warsaw CMS groups participate in searches for Heavy Stable Charged Particles (HSCPs) by looking for signatures of exotic particles with large masses, lifetimes of the order of a few nanoseconds and velocity significantly lower than the speed of light, as motivated by many BSM scenarios, including Split SUSY, Gauge Mediated SUSY Breaking models, and models with fourth generation leptons.  CMS has already produced stringent limits on gluino and stop R-hadrons, stau pair production and fourth generation fermions with electric charge 1 or 2.  Further improvements are expected from increased luminosity and development of new analysis techniques.

The ATLAS/ALICE Krakow groups
study BSM physics via photon-photon interactions in heavy-ion collisions, focusing on diphoton processes to search for axion-like particles (ALPs), and magnetic monopoles. ATLAS already set stringent ALP constraints (6–100 GeV) and excluded monopoles up to $m=120\,$GeV. Future ALICE/ATLAS measurements aim to enhance sensitivity to lower ALP masses and higher monopole charges.

The IFJ PAN group is involved in the MUonE experiment at SPS, which physics programme includes the direct search for exotic long-lived particles, such as dark photon or axion.

Polish groups have contributed to the development of the physics case, experimental programme and detector concepts for the future lepton colliders for over 20 years. Prospects for observing new physics signatures have always been of the primary interest. Many different BSM scenarios have been addressed, new studies were also initiated within the ECFA Higgs/Top/EW factory workshops series. Recent activities included searches for new exotic scalars, dark matter (DM) particles as well as heavy neutral leptons at ILC, CLIC and Muon Collider. Dedicated studies were also launched for long-lived particles (LLPs) searches at CLIC, ILC and FCC-ee.

For almost two decades Polish groups (UJ, CAMK more recently) have participated in the searches for direct interaction of DM particles with detectors based on noble gases, in particular on liquid argon. These experiments are complementary to the accelerator searches. The strong potential of the liquid argon technology makes it possible to push the sensitivity for DM detection several orders of magnitude beyond the current levels. Scientists from all the major groups working with argon (ArDM, DarkSide-50, DEAP-3600, and MiniCLEAN) have formed the Global Argon DARK Collaboration (GADMC) to explore the argon-based detectors potential.


\section*{IV~~~Flavour physics}

Flavour physics research in Poland is conducted across seven research units, spanning two major communities.
Polish experimental groups actively participate in the following ongoing flavour physics experiments: LHCb (the largest contribution) and CMS at the LHC at CERN, BESIII at the BEPCII accelerator in China,  and Belle II at the SuperKEKB accelerator in Japan. Among them, the involvement in detector construction and upgrades is concentrated  on LHCb and Belle-II. 

Studies of flavour physics proved to be a highly valuable tool in testing the Standard Model (SM) predictions as well as in searches for the BSM phenomena. 
In recent years, a wide programme of searching for new physics in the processes involving  $b$- and $c$-quarks yielded a number of interesting results in which the measured observables differ from the SM predictions by three or four standard deviations. The origin of these anomalies may be explained by some extensions of the SM as well as new physics hypotheses, for example enabling U-spin symmetry breaking. 
Thus, the priority of the Polish LHCb community is continuation of the aforementioned  new physics searches
using the full dataset collected at the HL-LHC. Such effort would yield a significant improvement in the experimental sensitivity to relevant observables with the associated tight constraints on the BSM models. 
The following topics are of particular interest for us: testing lepton flavour universality,
determining CPV phases in $B$ and direct CPV in charm sectors, 
CPT violation searches, rare decays and flavour-changing neutral
currents, searching for new hadronic states,
searches for inflaton and exotic processes with long-lived particles.
At the same time we plan to participate in the detector upgrade (Scintillating Fibre Tracker, RICH, Vertex Locator),
and data simulations. 

The general-purpose CMS experiment is complementary to LHCb in terms of kinematical coverage and luminosity reach. Thanks to its precise silicon detector and flexible trigger CMS has proven to effectively contribute to leading b-physics measurements and observations. The Polish CMS b-physics group has been recently established with key interests in rare decays.
The modernisation of CMS detector targeting LHC Phase-II, ensures that the CMS heavy flavour programme will be continued at highest luminosities. 

The primary goal of the BESIII is to investigate the properties of particle-antiparticle pairs produced in $e^+e^-$ collision above the $c\bar{c}$ threshold. The phenomenological methods of the study of
hadronic and semi-leptonic decays of baryons developed by Polish scientists are deployed in the BESIII analyses.

The Belle II experiment aims to perform precise measurements
of rare $B$ meson decays, as well as charm meson and tau lepton decays, and to probe for
effects beyond the Standard Model at the intensity frontier. The Krakow group is involved in analyses of semileptonic
and hadronic $B$ decays with complex experimental signatures and large missing energy, which
are challenging to study in hadronic collisions.

We are also involved in feasibility studies for future facilities. The NCNR BESIII group is investigating hyperon decays for the proposed STCF facility in China, with published research by Polish scientists serving as a foundation for these future investigations. The Krakow Belle II group is preparing for future upgrades of the SuperKEKB collider, working particularly on the implementation of a polarized electron beam and an ultra-high luminosity upgrade. In Europe, we are involved in discussions regarding the physics programme of the European Spallation Source (ESS) in Lund, Sweden. In particular, the NCNR, IFJ PAN and UJ groups are actively engaged in the preliminary planning of projects such as HIBEAM/NNBAR and REDTOP. These projects align with our broader interest in exploring fundamental symmetries and rare processes in particle physics.

\section*{V~~~Strong Interactions}

Experimental and theoretical physics of strong interactions (SI) is conducted in twelve research institutions distributed in seven cities in Poland. 
Our experimental groups involved in strong interaction physics participate in the following experiments at the large research laboratories all over the word:
    ALICE, ATLAS, LHCb at the LHC at CERN,
    NA61/SHINE, COMPASS, AMBER, MUonE at the SPS at CERN,
    STAR at RHIC at BNL, and
    HADES at FAIR at GSI.
    The largest contribution to one experiment is NA61/SHINE (61 participants).
Our groups are also planning to take part in the construction of detectors for new facilities like CBM and PANDA  at FAIR (GSI), EIC at BNL, and FCC at CERN.

Polish QCD theory and phenomenology research groups are focused on questions ranging from the behaviour of QCD in high-density environments to the effects of higher-order corrections and the extraction of parton distribution functions of protons and nuclei, also using lattice QCD. The studies cover low-$x$ QCD at high parton densities, quark-gluon plasma (QGP) studies, 3D imaging of hadron structure through exclusive and semi-inclusive processes, and entanglement in high-energy collisions, with strong ties to experiments such as ALICE, and new facilities FAIR, EIC, LHeC, and FCC.  
Polish theoretical groups are also working on the development of Monte Carlo generators, which are central to the research programme on strong interactions at the LHC and other facilities. The long-term plans address development of precision tools for higher-order calculations both in collinear and CGC factorization scheme and precise determination of PDF using lattice QCD.

\section*{VI~~~Neutrino and astroparticle physics}

Polish physicists contributed to neutrino studies for many years, participating in Super-Kamiokande and T2K experiments. 
Our involvement includes the detector work (testing, installations, upgrade) and physics analyses: oscillation studies, cross-section measurements, as well as the development of the NuWro event generator and theoretical models. Poland is also deeply involved in the future Hyper-Kamiokande (HK) project which targets a broad physics programme with unprecedented precision, including  neutrino oscillations, nucleon decays, and astrophysical neutrinos. Among others, we are responsible for the design and construction of a linear accelerator for calibration purposes, construction of the multi-PMT modules along with the corresponding electronics and underwater electronic modules for large-area PMTs. We plan to actively participate in the construction of the HK detector and the Intermediate Water Cherenkov Detector (IWCD), as well as in the calibration and data analysis.

Measurements of key importance for the reduction of uncertainties on the accelerator neutrino flux are performed by the NA61/SHINE experiment at CERN, with dedicated replicas of targets used by neutrino experiments, as well as thin targets. Future plans include construction of a multipurpose low energy beamline. Also, the unique measurements of hadron production and nuclear fragmentation from NA61/SHINE are directly related to understanding air showers induced by ultrahigh-energy cosmic rays, the interpretation of secondary and primary cosmic-ray fluxes, and searches for DM annihilation signals. Full and continuous support for the CERN North Area fixed-target programme is recommended in view of the continuation of these activities.

During the T2K upgrade and HK/IWCD preparations, the new detector prototypes and parts were tested at CERN on various particle beams, thanks to the CERN Neutrino Platform. We greatly appreciate CERN's support for neutrino experiments and strongly recommend continuation of the programme.

Neutrinoless double beta decay experiments are well recognized as the most powerful tool, not only to probe the Majorana nature of the neutrino and its mass (down to meV range), but also to investigate the neutrino mass hierarchy. The expansion of GERDA to LEGEND-200 and further to LEGEND-1000 provides unique discovery possibilities. Quasi-background-free operation and excellent energy resolution of the applied enriched HPGe detectors make the LEGEND experiment the one with the highest discovery potential in the field.

Polish groups are also involved in research with ultra-high-energy neutrinos with the aim to study the most extreme acceleration environments and mechanisms in the universe. This activity includes participation in KM3NeT, P-ONE and GRAND experiments with contributions to both hardware development, data analysis and software work.

The new challenges of particle physics require using complementary approaches and to design experiments which are able to detect different ``messengers''. Polish physicist participate in a range of such experiments, including Pierre Auger Observatory, CREDO, JEM-EUSO and POEMMA, Cherenkov Telescope Array Observatory and LIGO-Virgo-KAGRA Consortium, to which they contribute by designing and building equipment, developing software and performing analysis.

\section*{VII~~~Instrumentation}

Instrumentation must be a step ahead of the needs of the long-term European particle physics programme. The upgrade programmes and the incoming new facilities demand faster, radiation-harder, larger sensitive areas, and efficient, reproducible instruments to be at the forefront of the upcoming experiments. Instrumentation is pivoting from experiment-based R\&D toward specifically dedicated programmes which are experiment-agnostic. R\&D in instrumentation requires long-term planning, accumulation of know-how, technology, and training of staff over long periods. When resources are linked to short-term grants instead of stable R\&D programmes, advances and subsequent iterations of the technology become almost impossible. 

Following the recomendations of the ECFA Detector R\&D Roadmap, Detector Research
and Development Committee (DRDC) was formed at CERN and eight DRD collaborations
were established to support R\&D on different detector technologies,
bringing research groups to work together across
accelerators and experiments.
This strategy should also allow for broadening the community involved
in R\&D for instrumentation by incorporating expertise from other fields,
which may not be interested in joining the experimental collaborations. 
The ongoing DRD activities are leading toward the development of new
technologies that, eventually, will need to be adapted to the specific
needs of physics experiments at future facilities. Thus, close
collaboration between the R\&D community and experimental
working groups is still required.
More emphasis should be placed on the integration aspects of
instrumentation. The need for resources to create interdisciplinary
lab centres should be recognized at the state level, as well
as in EU funding programmes. 
Once the new flagship project is selected, the community working on
instrumentation and detector development would benefit from a unified
and coherent allocation of resources and efforts with the aim of
achieving critical mass in key projects.  

\section*{VIII~~~Theory}

Theoretical high-energy physics research in Poland focuses on several key areas. This includes (but is not limited to): (i) preparing Monte Carlo event generators for direct use by experimental collaborations, (ii) investigating extensions of the Standard Model and testing them against experimental data, (iii) performing precision calculations within the SM, with special focus on processes where signals of new physics could occur, (iv) studying QCD dynamics in bound states and in high-energy multipartonic processes, (v) studying strongly interacting matter at high temperatures and densities, and (vi) analyzing theoretical aspects of quantum field theories and string theory. 
The main research hubs are located in Kraków, Warszawa and Wrocław, with additional groups in Katowice, Kielce and Łódź, and smaller teams in Białystok, Poznań, Rzeszów, Szczecin, and Zielona Góra. 
The strength of Polish theoretical high-energy physics lies in its close connection to experiments.

Among the strategic research directions we name phenomenological studies and precision calculations within the Standard Model and its extensions as well as the development of Monte Carlo event generators, strongly represented in Poland. 
These activities need to be maintained in the future. 
It requires that funding agencies and experimental collaborations allocate appropriate resources for theoretical research along with experimental expenditures. 
This applies to the current European experiments like LHC as well as future ones like FCC which are and should remain the priority of theoretical research. The involvement of 
theoretical researchers in major projects outside Europe which rely on substantial international input and are in line with European scientific interests should also be supported. 
It is of utmost importance to create financial mechanisms to attract young researchers for whom we must compete with the hi-tech industry.

\section*{IX~~~Accelerator components and technologies}

Polish scientific institutions and industry have successfully participated in development, construction and commissioning of Big Science infrastructures for several decades. Key accelerator and detector components (Radio Frequency structures and systems, SC magnets, diagnostics and instrumentation, targets)  and technologies (cryogenics, electronics, radio frequency sources and others) designed, developed and manufactured in Poland have been deployed or are under construction in such machines as LHC, ESS, XFEL, FAIR, HL-LHC, SPIRAL2, HK and PIP II. The competences developed and maintained in Poland cover approximately 80\% of key components and technologies of superconducting Big Science accelerators.  Our most important national contributions to the accelerators comprise: cryogenic distribution systems design, production and installation; cryomodules installation and testing; RF electronics design and production; integration of superconducting (SC) magnets and cryogenics; risk analysis and testing of the accelerators subcomponents, beam dynamics studies and developments of novel collider instrumentation.

Free electron laser POLFEL is today a key project in Poland making use of accumulated Polish SC accelerator know-how and allowing the preservation of the acquired competences. No doubt, Poland is ready to participate actively in construction of new accelerators, such as FCC, linear \ee  or muon colliders, Energy Recovery Linacs, plasma accelerators and medical high intensity linear accelerators. The accumulated Polish supplies to CERN and other laboratories exceed 100 MEuro and are important factor of high-tech industry development and its entering international multibillion supply chains. Poland should maintain and actively support domestic Big Science projects, with POLFEL comprising in a small scale all key accelerator accelerators technologies. A dedicated strategic program should be established to support the scientific institutions and industry in R\&D efforts of accelerator technologies and components development.

Our community is also involved in the Gamma Factory project which aims to produce high-intensity gamma beams (outperforming the currently available gamma-ray sources by at least seven orders of magnitude) in laser beam interactions with highly ionised heavy atoms accelerated and stored at the SPS and LHC. These types of gamma rays can then be used to produce intense (surpassing the currently available sources by several orders of magnitude) beams of polarised electrons/positrons/muons, neutrinos, neutrons, and radioactive ions. This unique facility -- possible to be realised only at CERN -- opens up wide prospects for research in many fields of science, such as particle, nuclear, atomic and accelerator physics, astrophysics, and also creates great opportunities for many application solutions, e.g. in the field of nuclear energy and nuclear medicine. 

\section*{X~~~Computing}

Poland is home to several computing centres that house high-performance computers, with three of them, ACK Cyfronet AGH, NCBJ-CIS, and PSNC, serving as WLCG Tier 1 and Tier 2 resources for experiments conducted at the Large Hadron Collider (LHC). These centres also provide computing resources to local research groups. Additionally, local research groups operate small computer clusters for local calculation.
Currently, the allocation of computing resources for high-energy physics research in Poland is sufficient to meet existing obligations. However, given the ever-increasing demand for resources, particularly for LHC experiments, it is crucial to secure continued financial support. In the long term, this support should include an annual increase of around 20\% in resources for major computing centres. It is also important to support smaller local clusters, as these serve entire HEP community, including non-LHC experiments research groups.
Without these measures in place, Poland may face challenges in fulfilling its commitments to both LHC and non-LHC experiments. Looking further ahead, the resources accumulated during the LHC era should also be sufficient for preparatory work for future research initiatives such as FCC, new linear colliders or EIC, to name a few.

\section*{XI~~~Communication, Education and Outreach}
Whatever the details of the final recommendations of the Strategy Update, European particle physics community is facing perhaps the largest challenge of all time. The future big project will undoubtedly be the most complex single science endeavour ever, while at the same time being inherently disconnected from people's everyday experience. The lack of widespread well-targeted communication and education may have serious implications for both securing adequate level of future funding as well as ensuring steady flow of young enthusiastic people into science. With this in mind, efficient and compelling communication of the conducted research, long-term planning and benefits to society becomes paramount for the future of our field. Particle physics communication and outreach activities in all CERN Member States should be carried out in a coordinated manner and in line with the adopted strategy. The communication narrative should not be limited to research alone, but cover a wide range of topics including societal benefits, environmental impact and demonstrating that international scientific cooperation drives progress and peace, etc. In order to achieve these goals, more effort and resources need to be put into education, communication and outreach of particle physics. It is vital that these are coordinated at the European level with CERN acting as a hub and leader. Nonetheless, successful communication requires distributed efforts and dedicated resources allocated locally at national levels. We should strive to build a sense of ownership and responsibility for CERN within the Member States. This will provide support and facilitate the allocation of resources for the future projects.

\section*{XII~~~General conlusions and recommendations}

After very successful contributions to the construction of the LHC experiments, detector operation and data analysis, as well as ongoing activities focused on the HL-LHC detector upgrades, Polish teams are determined to contribute to the full exploitation of the physics potential of the LHC and the HL-LHC, including the studies of QCD and of flavour physics.
The community is also eager to participate in the design, construction and exploitation of the next energy frontier facility at CERN.
Previous updates of the European Strategy pointed to the \ee Higgs factory as the highest priority next large infrastructure. 
In recent years, CERN has firmly established the FCC as its preferred flagship project in the post-LHC era, and Polish teams are already involved in many FCC-ee (and FCC-hh) related activities, including in particular precision theoretical calculations.
The FCC-ee Higgs factory has numerous advantages.
It combines Higgs boson and top-quark studies with ultimate precision of the electroweak measurements at the $Z$-pole.
In addition, FCC-ee offers a unique possibility to probe the Higgs-electron coupling.
With four interaction points and four experiments, most of the CERN community can get involved in the project.
Preparations at CERN are well advanced, in particular arrangements for the FCC tunnel construction.
Furthermore, FCC-ee offers an excellent springboard to FCC-hh that provides direct exploration potential at uncharted energies, well beyond 10 TeV.

We find it very important that CERN continues to be the leading R\&D centre for accelerators and detectors, and the host for the future circular $e^+e^-$ collider. 
The above points to FCC-ee as the future best option for Europe and CERN, and we put it forward as the Polish recommendation for ESPPU. 
Possible construction of CEPC in China should not affect this decision.

The option of a linear collider as the future Higgs factory at CERN, with the new concept and running scenarios recently proposed, also offers competitive physics potential and should be taken as the second option.
For over 20 years, Polish groups have contributed to the development of the physics case, experimental programme and detector concepts for linear \ee colliders.
Higgs factory realized as a~Linear Collider Facility at CERN also has its advantages.
It has similar potential for Higgs and top quark studies 
as well as BSM constrains in the general SMEFT framework.
It also offers a flexible upgrade path. Depending on the final outcome of the HL-LHC and the results obtained at the initial $250\,$GeV stage, both luminosity and energy upgrades could be considered.
There are many well-founded models with diverse predictions, but BSM might look very different to what we expect and we need to be open to various possible scenarios, including different energy scales.
Diverse energy upgrade options (including through new innovative technologies) are a huge advantage, also allowing for much more flexibility in the face of ``unexpected'' scenarios.

It is crucial that diversity of particle physics research, of BSM search approaches in particular, is maintained at CERN and in the European countries. CERN should keep supporting non-collider and non-accelerator experiments, as well as the development of novel detector and  accelerator techniques, 
such as plasma acceleration, muon collider and Gamma Factory,
which should open up new perspectives for research in the future.
The Polish community strongly recommends building a new collider/detector at CERN to investigate QGP properties, hadron structure or non-linear QCD in the low-$x$ regime. 
We are equally strongly in favour of providing full and continuous support to fixed-target programmes at SPS, RHIC and FAIR which address unique and complementary aspects of QCD. Participation in major projects conducted outside Europe, such as the EIC being built in the US as well as neutrino and astroparticle experiments, should also be supported.

Education, communication and outreach play pivotal role in building trust and support for major scientific endeavours such as the future European particle physics project. To secure adequate level of future funding as well as ensuring steady flow of young enthusiastic people into science, more effort and resources need to be put into communication and outreach activities carried out across all CERN Member States, in a coordinated manner and in line with the adopted strategy.


\newpage

{\huge\bf Backup document}

\setcounter{page}{0}
\thispagestyle{empty}
\renewcommand{\thepage}{\arabic{page}}

\setcounter{section}{0}
\renewcommand{\thesection}{\arabic{section}}

\startcontents[sections]
\printcontents[sections]{l}{1}{\setcounter{tocdepth}{2}}

\section{Scope of the backup document}

The main aim of this document is to give a wider perspective of the current status and prospects of the Particle Physics (PP) research in Poland. 
It is to support the conclusions and recommendations included in the main document, prepared as Polish national input to the Update of the European Strategy for Particle Physics, initiated by the CERN Council in March 2024. 
The work was started by the Polish member of RECFA involving also the Polish delegation to the CERN Council as well as present and former ECFA members. 
Coordinating groups invited teams of experts, involving also ECRs, to prepare inputs to this document within different topical areas. 
The list of names of coordinating and topical groups is given at the end of this document.
After being presented at the community Town Hall meeting in January 2025, inputs were combined into the backup document while input summaries, also prepared by the expert teams, were used to form the main national input document. 
The document was then consulted with the HEP community in Poland at the second Town Hall meeting in March 2025. 

While the focus of this document is particle physics, there is a visible overlap nowadays between particle physics and astroparticle physics, as well as between particle and nuclear physics. It is not possible to rigorously separate astroparticle and nuclear physics from particle physics. In this document we tried to include all topics which seemed relevant for the future of particle physics research. 
Presented in Sec.~\ref{sec:overview} is an overview of the Particle Physics organisation and funding in Poland. Subsequent sections of the document present main activities, achievements and plans of the Polish HEP community in different topical areas.
For the conclusions and recommendations, please refer to the main document.

\section{Organisation and funding of Particle Physics in Poland}
\label{sec:overview}

%
%
Particle Physics research, including both experimental and theoretical activities, are being done in 18 research institutions - leading universities and research institutes - distributed in 8 centres in Poland. The two largest centres are Warsaw (7 units) and Kraków  (4 units).
The number of researchers in Polish PP counted in FTE is about 365 staff and 190 PhD and MSc Students. 
This is complemented by about 130 engineers and technicians directly involved in research.
List of academic units and research institutes involved in HEP activities in Poland is given in Tab~\ref{tab:units}. More details about Polish involvement in different groups of activities are given in Tab~\ref{tab:people}.


\begin{table}[tbp]

\renewcommand{\arraystretch}{1.2}

\centering
\begin{tabular}{|l|l|l|}
  \hline
  City & Unit & Acronym \\ \hline
%
%
Katowice & University of Silesia & UŚ \\ \hline
Kielce &  Jan Kochanowski University of Humanities and Sciences & UJK \\ \hline
Kraków &  AGH University of Kraków  & AGH \\
  & H. Niewodniczański Institute of Nuclear Physics & IFJ \\[-2mm]
  & \hfill Polish Academy of Sciences (PAS) & \\
  & Jagiellonian University & UJ \\
  & T. Kościuszko University of Technology & PK \\ \hline
Lublin & University of Maria Curie-Sklodowska & UMCS \\ \hline
Łódź  
     & Łódź University of Technology & PŁ \\ \hline
%
%
Toruń & Nicolaus Copernicus University & UMK \\ \hline
Warsaw &  Center for Theoretical Physics PAS & CFT \\
  &  Institute of Physics PAS  & IF\,PAN \\
 & Institute of Plasma Physics and Laser Microfusion & IFPiLM \\
  & National Centre for Nuclear Research (also in Świerk and Łódź) & NCBJ \\
  & Nicolaus Copernicus Astronomical Center  PAS& CAMK \\
   & University of Warsaw & UW \\
   & Warsaw University of Technology & PW \\ \hline
Wrocław &  University of Wroclaw & UWr \\
& Wrocław University of Technology & PWr \\ \hline
%
\end{tabular}

\caption{List of the institutions with HEP activities in Poland.
  The Polish acronyms given above are used further in this document. 
\label{tab:units}
}

\end{table}


\begin{table}[tbp]

\renewcommand{\arraystretch}{1.2}

\centering
\begin{tabular}{|l|c|c|c|c|}
  \hline
  Area of activity & ~~~Staff~~~  & Students & Support & ~~Total~~ \\ \hline\hline
  LHC experiments    &  90  &  70  & 60   &  220 \\
  Other exp. at CERN &  105 & 45  & 45  & 195  \\
  Exp. ouside CERN   &  55 & 20  & 25  & 100  \\ 
  Theory             &  115 &  55  &  - &  170 \\   \hline\hline
  Total             &  365 & 190  & 130  &   685\\  \hline
\end{tabular}

\caption{Estimated numbers of people involved in different HEP activities in Poland.
  Staff includes both permanent staff and post-doc positions. Students include both PhD and MSc student directly involved in research. Support includes engineers and technicians. 
\label{tab:people}
}

\end{table}

The funding  of PP in Poland comes from 4 main sources listed here in the order of importance: 
\begin{enumerate}[noitemsep,topsep=0pt]
    \item directly from the Ministry of Science and Higher Education (MNiSW) - mostly for the maintenance of the large scientific infrastructure in Europe, contributions to the large CERN experiments in particular,
    \item National Science Centre (NCN) - mostly for specific research activities through dedicated grants with funding period of 2 to 4 years, 
    \item National Centre for Research and Development (NCBiR) - focusing on technology
innovations with applications in industry,
\item National Agency for Academic Exchange (NAWA) - support for individuals coming to Poland in building their research groups, 
    \item Foundation for Polish Science (FNP) - dedicated grants for outstanding  individuals, 
    \item international cooperation agreements with other countries, and 
    outside funding agencies, including EU funding grants.
\end{enumerate}

As confirmed in the 2021 update of the European Strategy for Particle Physics (ESPP)~\cite{CERN-ESU-015} the successful completion of the Large Hadron Collider (LHC) program, including high-luminosity upgrade of the machine and detectors remains the focal point of European particle physics. 
Although Poland became a Member State only in 1991, we obtained the Observer status already in 1963.
After successful participation in construction and running of the LEP experiments, Polish physicists got involved from the very beginning in the design and construction of the four big LHC detectors. 
Hardware contributions included in particular: design and construction of the RPC readout and components of the L1 Trigger for CMS, contribution to the construction of the ATLAS Semiconductor Tracker, Transition Radiation Tracker and the ALFA detector, as well as to the development of the Time Projection Chambers and Photon Spectrometer for ALICE. More details on the Polish contributions to different experiment are included in Sec.~\ref{sec:inst_lhc}.
About 40\% of the research staff in experimental HEP in Poland is currently involved in operation and maintenance of the experiments at LHC, as well as in the data analysis and detector upgrades.
Polish teams are determined to contribute to the full exploitation of the physics potential of the LHC and the HL-LHC, including the study of flavour physics and the quark-gluon plasma.

Polish experimental groups are also very active at other CERN experiments. Experiment with largest participation is NA61/SHINE, where Polish physicists constitute about 50\% of the collaboration. Other teams contribute to COMPASS/AMBER and MUonE experiments at CERN SPS, as well as nuclear physics research at ISOLDE.
About 20\% of ongoing experimental activities are related to experiments outside CERN and outside Europe.
This includes in particular strong participation in the neutrino experiments (T2K, Super-Kamiokande, Hyper-Kamiokande), flavour physics experiments (BELLE\,II at SuperKEKB, Japan and BES\,III at BEPCII, China) as well as preparations for the experiments planned at Electron-Ion Collider (EIC) in the US. More details are given in the following sections.

The 2021 update of the European Strategy for Particle Physics~\cite{CERN-ESU-015} as well as the U.S. Community Study on the Future of Particle Physics (Snowmass '21)~\cite{Butler:2023glv} point to an $e^+e^-$ Higgs factory as the highest priority next large infractructure, to complete and deepen the successful Higgs boson measurements performed at the LHC and HL-LHC. 
In response to these recommendations the ECFA Higgs / top / electroweak Factory Study was set up to gather the experimental and theoretical communities involved in physics studies, experimental design and detector technologies towards future Higgs factories. Three working groups were established: on physics performance (WG1), physics analysis tools (WG2) and Detector R\&D (WG3). Furthermore, a number of 'Focus Topics' was defined by WG1 expert teams~\cite{deBlas:2024bmz}, where new studies were encouraged to reach full understanding of the physics potential of future \ee Higgs / top / EW factories. 

Polish research teams from Kraków, Katowice and Warsaw have replied to the ECFA invitation and made significant contribution to the study. In particular, the Kraków IFJ, UJ and Katowice University of Silesia groups work mainly on precision calculations for present and future collider studies (theory, Monte Carlo (MC) generators). University of Warsaw group, while also contributing to these topics, focuses mainly on the prospects for direct observation and identification of BSM phenomena covering a wide range of possible scenarios. 
Activities of Polish groups resulted in multiple contributions which were presented at the ECFA workshops and included in the final report of the ECFA study~\cite{ECFAreport}. The main results are also summarized in the following sections of this report.

\section{Electroweak and Higgs boson physics}

\subsection{Experimental involvement} 

Physicists from Kraków and Warsaw are deeply involved in studies of heavy weak and Higgs bosons with LHC data collected by the ATLAS (Kraków) and CMS (Warsaw) experiments. These studies can be grouped as follows.

\subsubsection*{The $Z$ boson measurements}
The Kraków group was involved in a series of precise measurements of the $Z$ boson in the ATLAS experiment. 
The group measured the transverse momentum distribution of $Z$ in the data collected at an energy of 8~TeV~\cite{Aad_Z_2024} and interpreted these results for the measurement of the cross section and the differential distributions of the $Z\to ll \gamma$ and $Z\to ll \gamma\gamma$ processes~\cite{Aad_Zll_2024}. 
Finally, the measurement of the effective leptonic weak mixing angle using lepton pairs from $Z$-boson decays was performed~\cite{ATLAS-CONF-2018-037}.

\subsubsection*{Probing vector boson scattering}
Measurement of cross-section for various massive vector-boson scattering configurations ($VV\to VV$) is one of promising windows to probe physics beyond the Standard Model (SM)~\cite{Buarque_Franzosi_2022}. These studies are motivated by precise cancellations of various contributions to the $VV\to VV$ scattering processes built in the SM which can be modified by its extensions. These measurements involve precise testing of SM predictions regarding the kinematic distributions of the $W$ and $Z$ bosons in the final state. The results are then interpreted using the effective field theory (EFT) approach, i.e. by extending the SM Lagrangian with all permissible operators describing the interactions of the $W$ and $Z$ bosons in the presence of exotic interactions not accounted for in the SM. The CMS Warsaw group is involved in phenomenological (together with the theory groups from University of Warsaw and Institute of Nuclear Physics, IFJ Kraków)~\cite{Buarque_Franzosi_2022, Dedes_2021, gallinaro2020standardmodelvectorboson, Chaudhary:2021zej,Szleper:2020kcv,Kalinowski:2018oxd} as well as experimental studies on the same-sign~$WW$ scattering, which is expected to be the most experimentally accessible. A significant contribution to the analysis of the Run~3 LHC data, which will result in an extension of present exclusion limits with Run~2 data~\cite{CMS:2020gfh, CMS:2020etf}, is expected.

\subsubsection*{Single Higgs boson phenomenology and measurements}
The ATLAS group was involved in the measurement of Higgs boson production cross-sections using its decays to a pair of leptonically decaying $W$ bosons, $H\to WW^{*}\to e\nu \mu \nu$ at $\sqrt{s}=13$~TeV.
The cross-sections for Higgs production through gluon fusion and vector boson fusion~\cite{atlas-higg-2016-07} were measured, followed by cross sections via its associated production with vector bosons~\cite{Collaboration:2838206}.
Additionally, the  measurement of  Higgs boson couplings to polarised $W$ and $Z$ bosons in the vector boson fusion production and  $H\to WW^{*}$ decays was performed~\cite{atlas_HWWspinCP}
using information on measured cross sections and the distribution of azimuthal angles between forward jets.

A simulation tool for measuring the CP state of the  Higgs boson in its decays to $\tau$ leptons, {\sf TauSpinner}, has been developed~\cite{Przedzinski:2014pla}. 
It utilises the transverse spin correlations corresponding to any mixture of CP-even (scalar) and CP-odd (pseudoscalar) states and is now widely used by both ATLAS and CMS experiments.
Phenomenology studies in optimizing the Higgs boson CP measurement in $H\to \tau\tau$ decays at the LHC by applying  machine learning techniques~\cite{Jozefowicz:2016kvz, Lasocha:2020ctd} have been completed.
Measurement of CP parity in the decay of the Higgs particle into a pair of tau leptons has been performed~\cite{Aad_CPHtautau2023}, within the ATLAS experiment, confirming its CP-even nature.

Study of the Higgs boson decays into $\tau$ leptons pairs have also been performed by the Warsaw group with the CMS detector~\cite{CMS:2017zyp,CMS:2021gxc,CMS:2022kdi,CMS:2024jbe}. Recently, the group contributed to the study of the CP structure of the Higgs and $\tau$ lepton Yukawa coupling was performed with the CMS detector by the Warsaw group. The measurement excluded pure CP-odd nature of the coupling at 3.0 standard deviations and the effective mixing angle between CP-even and CP-odd couplings was found to be $-1\pm 19^{\circ}$ at 68\% CL~\cite{CMS:2021sdq}. The study is continued with Run~3 data with improved analysis techniques.

Finally, the $\tau\tau$ final state have been used to search for additional Higgs bosons predicted by extensions of the SM (with both ATLAS and CMS); no signal was found and exclusion limits were set~\cite{ATLAS:2020zms,CMS:2022goy}.

\subsubsection*{Higgs boson pair searches}

Members of the Kraków ATLAS group have been active in analyses targeting several final states: $bb\gamma\gamma$~\cite{ATLAS-CONF-2016-004} and multileptonic without $b$ quarks  (involving $\tau$ leptons and $W$ bosons)~\cite{Aad_2024_HHmultilepton}. 
Moreover, they have been involved in preparing statistical combinations of di-Higgs searches in both Run~1~\cite{Aad:2015xja} and early Run~2~\cite{Aad:2900946}.
These works focused on searches for di-Higgs production in the SM and beyond, and in measuring Higgs trilinear self-coupling, crucial for determining the shape of the Higgs potential.

The group was also involved in early projections of the sensitivity to Higgs pair production in the $b\bar{b}\gamma\gamma$ final state with the ATLAS detector at the High Luminosity LHC (HL-LHC)~\cite{ATL-PHYS-PUB-2014-019}, prepared for the ECFA 2014 Conference. 

\subsection{Electroweak calculations for LHC and future colliders}

The Kraków theoretical groups are working on key calculations and tools needed for the (HL)-LHC electroweak program, such as the Herwig~\cite{Bellm:2015jjp} and Sherpa~\cite{Sherpa:2024mfk} general-purpose Monte Carlo generators or the WINHAC generator~\cite{Placzek:2003zg}, which was used in a recent measurement of the mass of the $W$ boson at the LHC. Regarding other experiments currently in operation, versions of KKMC-ee~\cite{Jadach:2022mbe} and Tauola~\cite{Jadach:1993hs} are
used in Belle II collaboration and offer the gate for improvements (fits to Belle data) of tau lepton decay form-factors.

Concerning future colliders, Kraków IFJ, UJ groups and Katowice theory group focus on the Standard Model calculations for the future Higgs factory, targeting requirements of the Future \ee Circular Collider (FCC-ee) in particular. 
With $Z$-pole, $WW$ and top threshold running options, the physics program of future \ee colliders extends significantly beyond the study of the Higgs boson, especially opening the discovery potential via precision measurements. Circular colliders offer the unique opportunity of delivering very high luminosity at the $Z$ pole. 
Thanks to tremendous technical progress, FCC-ee will be 4-5 orders of magnitude more efficient than LEP. Collecting nearly $10^{13}$ Z bosons at $e^+e^-$ resonance collisions (thus also called the TeraZ $e^+e^-$ mode), the proposed CERN circular collider will improve the precision of determination of basic electroweak pseudo-observables of the Standard Model by one to two orders of magnitude~\cite{Blondel:2018mad}. This will define the background for direct and indirect BSM searches. 
Circular colliders make also possible an excellent measurement of the centre of mass energy~\cite{Blondel:2019jmp}, by resonant depolarisation of pilot bunches, which can only be done at circular colliders.
Precise determination of the centre of mass energy  affects sensitivity of observables measurements at the Z pole and WW threshold ($\Gamma_Z, m_Z, m_W$)~\cite{EWPrecisionWshop}.  
For all other measurements, the FCC-ee TeraZ run offers an extremely sensitive program of electroweak, QCD, and flavour physics, as well as numerous searches for rare processes~\cite{Blondel:2019yqr,blondel_2024_jg108-xp787}.

To meet the precision goals of the discussed future high statistics, significant advances in calculations of higher-order radiative corrections and MC generators will be needed~\cite{Blondel:2018mad,EWPrecisionWshop,Freitas:2019bre}. For instance,  electroweak NNLO corrections for various pair production processes ($e^+e^- \to \rm f\bar{f}$, $e^+e^- \to \rm W^+W^-$, $e^+e^- \to \rm ZH$) are needed, as well as MC tools for the simulation of multiple photon radiation beyond the leading-logarithmic approximation. Moreover, three-loop corrections in the full SM and leading four-loop corrections will be required to interpret precision measurements at the Z pole.  
Working in this direction, the Katowice theory group continues, after completing SM precision calculation of Z-pole EWPOs at the two-loop level~\cite{Dubovyk:2019szj}, the efforts in the computation of the 3-loop missed contributions $\alpha\as^2$, $\alpha\as^3$, $\alpha^2\as$, $\alpha^3$ to the $Z$-pole EWPOs and SM parameters such as $Z$-boson partial and total decay widths, forward-backward asymmetry, weak mixing angle(s), muon decay $\Delta r$. The group also develops methods and tools needed for analytical and numerical multiloop calculations~\cite{Dubovyk:2022obc}. 
The plans of Kraków groups are connected with further development of the Monte Carlo event generator {\sf KKMCee}~\cite{Jadach:2022mbe} (implementing new processes $e^+e^- \rightarrow HZ \rightarrow 4f$ and $e^+e^- \rightarrow \gamma\gamma$, merging {\sf KKMCee} with EW libraries). Both teams plan to implement the N$^3$LO QED matrix element in the YFS scheme, improve the {\sf BHLUMI} and {\sf BHWIDE} MC event generators for Bhabha scattering, develop analytical and numerical methods and tools for efficient evaluation of Feynman integrals and MC generators.
Focusing more on the TeV scale lepton colliders \cite{LinearColliderVision:2025hlt}, Warsaw group contributes to the development of the {\sf WHIZARD} event generator~\cite{Kilian:2007gr,Moretti:2001zz}, implementation of the electroweak PDFs in particular~\cite{Reuter:2024dvz}, and different studies of radiative processes \cite{Kalinowski:2020lhp,Mekala:2024hdv,Mekala:2025rlk}. 


\section{Beyond Standard Model and Dark Matter searches}

\subsection{Experiments at the LHC}

The ATLAS/ALICE Kraków groups (AGH, IFJ PAN) are engaged in performing BSM searches using heavy ion collisions, by exploring photon-photon interactions. These interactions occur due to extremely strong electromagnetic fields produced by the colliding lead nuclei (ultraperipheral collisions, UPCs). In particular, processes with diphoton final states provide a very clean channel to search for axion-like particles (ALPs), using the $\gamma\gamma \to \,$ALP$\, \to \gamma \gamma$ reaction. 
This is already demonstrated by ATLAS~\cite{ATLAS:2020hii}, by providing the most stringent constraints in the search for ALP signals in the mass range 6–100 GeV. With larger datasets, future measurements by ATLAS/ALICE have a potential to further improve the search sensitivity, with ALICE covering even lower ALP masses (below 6 GeV).
Another process of interest is the production of monopole pairs in UPC. This can be theoretically described by the dual Schwinger process, avoiding problems with large coupling constant of magnetic monopoles. ATLAS has published a recent search based on LHC Run 3 data, excluding masses of up to 120 GeV for monopoles with a single Dirac charge~\cite{ATLAS:2024nzp}. Future planned measurements can extend this exclusion region further, also by exploring higher values of monopole magnetic charge.

Kraków ATLAS groups are actively involved in searches for beyond the Standard Model (BSM) phenomena in the Higgs sector. Various non-minimal Higgs scenarios, such as Two-Higgs-Doublet Models (2HDM) or models containing Higgs triplets predict existence of heavy neutral pseudoscalar (A) and scalar (H) states as well as charged Higgs bosons, H$^\pm$, in addition to the already discovered 125 GeV light scalar (h). BSM Higgs bosons could additionally manifest themselves through a substantially increased cross-section for h pair production. Kraków activities concentrate on scenarios where scalars decay to either $\tau$ leptons or b quarks with special expertise in tau identification, background modeling, machine learning-based discriminants and final statistical analysis. A model independent search for H$^\pm$ produced in top-quark decays or in association with a top quark and decaying to a $\tau$ lepton and a neutrino was carried out with 36.1\,fb$^{-1}$ of proton-proton collision data at $\sqrt{s}$=13\,TeV~\cite{ATLAS:2018gfm} and is currently upgraded to the full Run 2 statistics (140\,fb$^{-1}$)~\cite{ATLAS:2024hya}. A search for heavy (pseudo)scalar H/A decaying into a $\tau^+\tau^-$ pair was performed using the full Run 2 data, with H/A produced either via the gluon fusion or in association with a b-quark~\cite{ATLAS:2020zms}. Currently the analysis is being extended to lower masses (80-200 GeV), especially to cover the experimentally difficult range of masses around and below 125 GeV. Another analysis focused on heavy neutral Higgs bosons produced in association with one or two b-quarks and decaying into b-quark pairs. The search was performed in the mass range 450–1400 GeV using 27.8\,fb$^{-1}$ of $\sqrt{s}$=13\,TeV proton-proton collision data~\cite{ATLAS:2019tpq} and is now being updated to full Run 2 statistics. A search for non-resonant Higgs boson pair production in final states with leptons using 140\,fb$^{-1}$ of $\sqrt{s}$=13\,TeV data has recently been published~\cite{ATLAS:2024lhu} and provides constraints on anomalous trilinear Higgs coupling modifier $\kappa_\lambda$. 
A combinbation of searches for Higgs pair production in early Run 2~\cite{ATLAS:2019qdc} combines six final states with b-quarks, W bosons, tau-leptons and photoins to set limits on the production of narrow spin-0 resonances and spin-2 Kaluza-Klein Randall-Sundrum gravitons. Additionally, BSM effects were looked for in the Higgs boson couplings to polarised W and Z bosons~\cite{ATLAS:2021pkb}.
All analysis activities are being continued to cover data accumulated in LHC Run 3 as well as extend to the HL-LHC period.

The NCBJ Warsaw group of the CMS experiment is deeply involved in searches for new Heavy Stable Charged Particles (HSCPs).  New particles, with masses greater than about 100 GeV, lifetimes of the order of a few ns and velocity significantly smaller than the speed of light are expected in many BSM scenarios, including Split Supersymmetry, Gauge Mediated Susy Breaking models and models with fourth generation leptons.  The signature consists of a single isolated high-pT track with large ionization energy loss in the silicon tracker, where the mass of the traversing particle can be deduced from dE/dx of the track.  In addition, if the particle lives long enough to survive to the outer detector (muon chambers), time-of-flight techniques can be applied.  Searches for HSCPs were conducted on data collected in 2015 (2.5 fb$^{-1}$)~\cite{CMS:2016nhn}, 2016 (12.9 fb$^{-1}$), and most recently 2017-2018, the latter corresponding to a total integrated luminosity of 101\,fb$^{-1}$)~\cite{CMS:2024nhn}.  The results revealed no excess of events over the SM background and placed significant limits on the pair production of supersymmetric particles, namely gluinos, top squarks, tau sleptons, and of Drell-Yan pair production of fourth generation $\tau'$ leptons.  These searches will clearly benefit from increased statistics as expected from Run 3 data and the future HL-LHC phase. Foreseen is a continuation of this analysis, with significant improvements in the scope of the search and in the employed analysis techniques.  In particular, the Warsaw group has been involved in developing time-of-flight techniques to measure the particle's velocity.  Other considered improvements include better dE/dx discriminators using Machine Learning.  The analysis will be also extended to consider decaying particles (limits depending on $c\tau$) and multiply charged particles.

The engagement of the LHCb group from IFJ PAN in Kraków in the BSM considers the searches for experimental signatures with heavy particle jets and exotic particles decaying in displaced secondary vertices, such as the measurement of the $Z^0(\to\mu^+\mu^-)\,b\,\bar{b}$ production cross section. Owing to the presence of heavy quarks, it is sensitive to a variety of BSM phenomena. In particular, the higher order corrections induced by exotic particles appearing in the quantum loops or the production enhancement due to similar experimental signatures (e.g. $b^\prime \to Z^0\,b$ or $Z^0\,Z^\prime \to Z^0\, b\,\bar{b}$, where $b^\prime$ is a hypothetical 4th generation quark). Both would manifest as the deviation of a measured production cross section from its value predicted by the Standard Model. Furthermore, $Z^\prime$ and $b^\prime$ are expected to be rather massive and decay into boosted final states in the forward kinematic region, available primarily to the LHCb experiment. Another study conducted by the LHCb group from IFJ PAN in Kraków is related to the searches for exotic BSM decays with displaced vertex signature, such as exotic particles predicted by the Hidden Valley models~\cite{Strassler:2006im,Strassler:2006ri}, where a hidden sector is taken to have a QCD-like structure and the communicator to be the Standard Model Higgs boson. The Higgs boson is then assumed to decay into a pair of the so-called hidden valley pions, which subsequently decay exclusively into a pair of b-quarks each. Such analyses~\cite{LHCb:2014jgs,LHCb:2017xxn,LHCb:2021dyu}  are supposed to be continued for the detector upgrades for Run 3 and beyond.

Physicists from the University of Warsaw have been involved on theory/phenomenology level in the searches for exotic highly-ionizing particles by the MoEDAL collaboration~\cite{MoEDAL:2021mpi}. Plans are to continue long-lived charged particle searches with the MoEDAL and MoEDAL-MAPP detectors. In particular, estimate the sensitivities of MoEDAL-MAPP to theoretical models predicting long-lived millicharged particles and compare them with other phenomenological constraints.

\subsection{Other ongoing experiments}

The DarkSide-20k (DS-20k) detector has been designed to search for direct interactions of DM particles (WIMPS) applying liquid argon (LAr, dual phase TPC) as a target. 20 t of argon depleted in Ar-39 by a factor of 1400 will be used as the fiducial volume. The DS-20k detector will have ultra-low backgrounds and the ability to measure it in situ, resulting in an expected sensitivity to WIMP-nucleon cross sections of $6.3\cdot 10^{-48}$\,cm$^2$ for 1\,TeV/c$^2$ WIMPs following a 10 yr run with a total fiducial volume exposure of 200 t$\cdot$yr~\cite{DarkSide-20k:2017zyg,DarkSide-20:DS2,DarkSide-20k:2023fbh}. This projected sensitivity is significantly better than currently published results above 1 TeV/c$^2$ and covers a large fraction of the parameter space currently preferred by supersymmetric models. During the 200 t$\cdot$yr exposure single background events are expected from the coherent scattering of atmospheric neutrinos, making DS-20k the first ever direct dark matter detection experiment to reach this milestone. DS-20k experiment is foreseen to begin taking data in 2026 and will either detect WIMP dark matter or exclude a large fraction of favored WIMP parameter space. DS-20k searches are complementary to collider searches: the observations of WIMPs with masses up to about 1 TeV/c$^2$ is a major objective of the experimental program at the High Luminosity Large Hadron Collider. Future high energy colliders like the FCC-hh (Future Circular Collider) will be able to extend these searches up to the 10 TeV/c$^2$ mass range. 
Given the potential reach of an argon-based detector, scientists from all of the major groups currently using LAr to search for dark matter, including DS-20k, ArDM, DS-50, DEAP-3600 and MiniCLEAN, have joined to form the Global Argon Dark Matter Collaboration (GADMC) with a goal of building a series of future experiments that maximally exploit the advantages of LAr as a detector target. Polish groups are actively participating in the collaboration activities: UJ is involved in various activities related to background reduction of the detector; CAMK contributes to the design of novel LAr scintillation light detectors based on SiPMs; PŁ participates in the studies of LAr scintillation light properties; PW is involved in development of DAQ and data processing systems.

A novel experiment at SPS, called MUonE~\cite{Abbiendi_2017,Abbiendi:2677471}, has been proposed to measure the hadronic component of the running electromagnetic coupling in a momentum transfer region relevant for the calculation of the muon g-2 (see also Sec. \ref{sec:si1}). However, the design of MUonE detector and clean experimental environment allow also for an effective direct search for exotic long-lived particles decaying outside of the target volume, such as a dark photon or axion~\cite{Galon:2022xcl}, being produced in the scattering and decaying leptonically or hadronically after travelling some distance from the target. The MUonE group from IFJ PAN in Kraków is actively engaged in the preparation of such analyses, including simulation, analysis strategy and appropriate tools, as well as redesigning of the detector in order to allow the sensitivity to various masses and lifetimes of the exotic particles~\cite{Abbiendi:2021xsh}. The MUonE group from IFJ PAN is fully responsible for the central MUonE software based on FairRoot, including algorithms for all the crucial stages of experimental data analysis, i.e. event reconstruction, detector simulation, alignment and offline selection.

\subsection{Future projects}

Polish groups have contributed to the development of the physics case, experimental program and detector concepts for the future lepton colliders for over 20 years. Prospects for observing “new physics” signatures have always been of the primary interest. Many different BSM scenarios have been addressed in close collaboration between theoretical and experimental groups. Many of the corresponding contributions were included in the ECFA study report~\cite{ECFAreport} and to the Linear Collider Vision White Paper \cite{LinearColliderVision:2025hlt} for the ESPP update.

One of the first models extensively studied at the University of Warsaw was the Inert Doublet Model (IDM), one of the simplest extensions of the SM, which can provide a dark matter candidate. A large set of benchmark scenarios was first selected, consistent with all current constraints on direct detection, relic density of dark matter, as well as collider and low-energy limits, covering the large variety of collider signatures that can result from the model~\cite{Kalinowski:2018ylg,Kalinowski:2020rmb}. Prospects of discovering the IDM  scalars at CLIC were then studied in two signal processes:  pair-production of charged inert scalars ($e^+e^-\to H^+H^-$) and pair-production of the neutral ones ($e^+e^-\to A\, H$), followed by decays of charged scalars $H^\pm$ and neutral scalars A into leptonic final states and missing transverse energy~\cite{Kalinowski:2018kdn,Zarnecki:2020swm}. A leptonic signature allows for very efficient background suppression. However, the CLIC sensitivity to IDM scalar pair-production in this channel is limited to masses below about 500 GeV due to the small branching ratio. IDM scalars with higher masses can still be discovered at high energy stages of CLIC (1.5 TeV and 3 TeV) in the semi-leptonic final state~\cite{Klamka:2022ukx}. 

High energy \ee colliders offer unique possibility for the most general dark matter (DM) search based on the mono-photon signature. Analysis of the energy spectrum and angular distributions of photons from the initial state radiation can be used to search for hard processes with invisible final state production. The procedure developed at the University of Warsaw allowed for consistent, reliable simulation of mono-photon events in the Whizard framework, for both BSM signal and SM background processes, based on merging the ME calculations with the lepton ISR structure function~\cite{Kalinowski:2020lhp}. Sensitivity to light DM production was studied as a function of the mediator mass and width based on the expected distributions of the reconstructed mono-photon events for experiments at the International Linear Collider (ILC) and Compact Linear Collider (CLIC)~\cite{Kalinowski:2021tyr,mttd2021}. For light mediators, coupling limits derived from the mono-photon analysis are more stringent than those expected from direct resonance searches in decay channels to SM particles.

Models with heavy neutral leptons (HNL) of Dirac or Majorana nature address many open problems of the Standard Model such as the neutrino masses, baryon asymmetry in the Universe or the nature of the dark matter (DM). For scenarios when the HNL production mechanism is via the weak force, lepton colliders seem to be the most suitable devices for their searches. Prospects for observation of heavy HNL production were studied for experiments at the ILC, CLIC and the Muon Collider (MuC)~\cite{Mekala:2022cmm,Mekala:2023diu}. The limits for the future lepton colliders, extending down to the coupling values of $10^{-7}$ to $10^{-6}$, are orders of magnitude stronger than the expected reach of hadron colliders. With use of Machine Learning methods, one may also efficiently discriminate between Dirac and Majorana nature of the heavy neutrino(s) simultaneously with their discovery~\cite{Mekala:2023kzo}. 

While the physics program for the future \ee Higgs factory focuses on measurements of the 125 GeV Higgs boson, production of new exotic light scalars is still not excluded by the existing experimental data, provided their coupling to the gauge bosons is sufficiently suppressed. Sizable production cross sections for new scalars can also coincide with non-standard decay patterns, so a range of decay channels should be considered. Little studied in the past, this was selected as one of the ``focus topics'' to be studied within the ECFA Higgs/Top/EW factory workshops series~\cite{deBlas:2024bmz}. The University of Warsaw group has made a leading contribution to this topic addressing experimental prospects for different decay channels of the exotic scalar~\cite{Zarnecki:2024xdy,Brudnowski:2024iiu}. Highest sensitivity to the light scalar signal at the 250 GeV Higgs factory (limits about an order of magnitude stronger than from the decay independent approach)  is expected in the tau pair decay channel. 

In recent years, more and more attention is put to searches for long-lived particles (LLPs)  as an exciting alternative and complement to traditional searches focusing on heavy new states with prompt decays. Independently of the BSM model considered, LLPs searches are essentially signature-driven. A generic case of neutral LLPs decaying to any two (or more) charged particles forming the displaced vertex was considered for the ILD detector at the 250 GeV ILC~\cite{Klamka:2023kmi,Klamka:2024gvd}. Two classes benchmark scenarios were selected as challenging from an experimental perspective: pair-production of heavy neutral scalars, A and H, where the former is the LLP and the latter is stable and escapes the detector undetected, and production of a very light and highly boosted LLP with strongly collimated final-state tracks. The expected limits on the LLP production cross section were extracted for a wide range of the LLP proper lifetimes corresponding to $c\tau$ from 0.1\,mm to 10\,km. The displaced vertex reconstruction procedure was then applied to the scenario with 125\,GeV Higgs boson decaying to two neutral LLPs. Expected limits on the Higgs branching ratio in the considered channel go down to about $10^{-4}$. 

The engagement of the FCC group from IFJ PAN in Kraków includes exotic long-lived particle searches, i.e. the FCC-ee sensitivity to the Hidden Valley particles, with the Standard Model Higgs boson as a communicator, decaying into a pair of exotic Hidden Valley pions, which are assumed to subsequently decay exclusively into pairs of Standard Model b-quarks, but the results can be scaled as well to account for other channels.

Concerning the non-accelerator Dark Matter searches with detectors based on LAr the  ultimate objective, towards the end of the next decade, would be the construction of the ARGO detector. Its fiducial mass of 300 t of underground argon (depleted with Ar-39) would be sufficiently big to push the sensitivity to the region where neutrino background will be a limitation in detectors without directional capability~\cite{DarkSide-20:DS2}. The WIMP detection sensitivity will only be limited by systematic uncertainties in nuclear recoil background from coherent neutrino scattering of atmospheric neutrinos. The strong electron recoil rejection will eliminate background from solar neutrinos and some residual internal backgrounds such as radon. The needed amount of underground argon would be produced within 6 years. SNOLAB in Canada is considered as a potential site for this detector, which would also enable the observation of solar neutrino sources (CNO, pep)~\cite{Franco:2015pha}. Groups from UJ, CAMK and PW are strongly involved in this project.


\section{Flavour physics}

\subsection{The experiments at LHC -- Europe}

\subsubsection*{LHCb}

The LHCb experiment has the world's largest sample of beauty and charm hadrons and is ideally suited for new physics searches in CPV studies, rare and exotic decays. Studies in beauty sector showed CPV up to 75\%  (the largest value not observed till now). Studies in the charm sector are reaching theory precision $10^{-4}$ allowing for sensitive tests in a unique rapidity range.  From the start of LHC till now, the LHCb has discovered about 60 new exotic particles from about 75 identified by all experiments at LHC. 

The importance of indirect measurements, such as those carried out at the LHCb experiment, has increased significantly our construction and understanding of the SM. In recent years, a wide program of searching for new physics in the $b$- and $c$-hadrons provided a number of interesting measurements. A couple results differ from the SM predictions by 3 or 4 standard deviations. The origin of these anomalies may be explained by some extensions of the SM as well as new physics hypotheses, for example enabling U-spin symmetry breaking. 
In addition to physics analyses, the Polish groups have contributions in 
detector prototyping and construction, simulation of the detector response and optimization of the detector subsystems, development of trigger algorithms and physics analysis tools with advanced machine learning techniques.
The priority of the Polish LHCb groups is to continue the new physics searches in very precise measurements of the SM phenomena 
using large statistics of data 
collected after the 2023 detector upgrade,
with the sensitivities corresponding to theory constraints or even better.
The main topics of the planned Polish involvement include: testing Lepton Flavour Universality (LFU), studies of CP violation in Quark Sectors (CPV phases in $B$ and direct CPV in charm), searches for CPT violation (CPTV), rare decays and Flavour-Changing Neutral Currents (FCNCs), searches for new hadronic states (hadron spectroscopy) and BSM physics: inflaton and exotic processes with long-lived particles.
Participation in the hardware development and construction of the future detector in next upgrade phase will include detector contribution (Scintillating Fibre Tracker, RICH, Vertex Locator) as well as data simulations.

\subsubsection*{CMS}

The CMS experiment at the LHC is a general-purpose experiment with sensitivity to charged tracks in the pseudo-rapidity region $|\eta|<2.5$
enabling studies in the kinematic region complementary to LHCb.
A high-precision silicon pixel detector guarantees precise reconstruction of particle tracks through the detector at the full luminosity of the LHC accelerator, providing a huge amount of data on the physics of heavy flavours.
At the same time, it is possible to reconstruct particle tracks and the positions of primary and secondary vertices with sufficient precision to observe weak decays, in particular decays of particles containing $b$ and $c$ quarks.

Thanks to this, CMS effectively contributes to the study of the production of heavy quarks and quarkonia, the exploration of CP violation,
study of decays of mesons and baryons containing heavy quarks in exclusive channels, and searches for rare decays, such as: $B^0/B_s^0 \rightarrow\mu^+ \mu^- $ or for exotic states.
Simulations show that, without loss of precision, research can continue the in luminosity expected in phase II of the LHC, which will further increase statistics and allow for improved precision of measurements and possible discoveries.
The CMS Polish b-physics group has been recently established with key interests in rare decays.

\subsection{The BESIII experiment in China}

NCBJ is involved in the  BESIII experiment \cite{BESIII:2009aa} at the BEPCII electron--positron collider \cite{Ye:1987nh} in Beijing. The primary goal is to investigate the properties of particle-antiparticle pairs produced from these collisions. The BESIII experiment gathers data in the energy range of 2 GeV to 4.9 GeV, which facilitates the production of hyperons and charmed baryons and tests of  CP symmetry. To date, CP violation  has only been observed in meson decays and no  observation evidence of the effect in baryon decays.

Polish physicists focus on the study of hyperon-antihyperon production and the research of two-body and semi-leptonic decays of strange baryons. The modular approach to analyzing the angular distributions of baryon-antibaryon pairs, developed by the Polish group \cite{Salone:2022lpt,Batozskaya:2023rek}, is widely utilized in many analyses within the BESIII collaboration.
Following the upgrade in 2024, the collaboration plans to collect a significant sample of baryon-antibaryon pairs containing a charmed quark. The phenomenological methods of the study of hadronic and semi-leptonic decays of baryons developed by Polish scientists will be employed in the analysis of these new data samples. 

Currently, only the NCBJ group participates in BESIII, but there is growing interest from other Polish institutes to increase Poland's contribution.

\subsection{The Belle II experiment at the SuperKEKB in Japan}

The Belle II experiment, a second-generation B factory, 
aims to perform precise measurements of rare B meson decays, as well as charm meson and tau lepton decays, and to probe for effects beyond the Standard Model (BSM) at the intensity frontier.
During Run 1 (2019–2022), the experiment accumulated an integrated luminosity of approximately 400 fb$^{-1}$.
A major upgrade to the accelerator and detector systems was carried out in mid-2022, with the experiment resuming operations in the fall of 2023 for Run 2. By December 2024, the experiment had already collected 150 fb$^{-1}$ and set a new peak luminosity record of 5.1 $\times$ 10$^{34}$ cm$^{-2}$s$^{-1}$. The goal for Run 2 is to collect up to 10 ab$^{-1}$ over the next four years. A substantial upgrade (Long Shutdown 2) is planned after Run 2 to enable the collection of tens of ab$^{-1}$. The ultimate target is 50 ab$^{-1}$ by the early years of the next decade. 

The Polish group from the Henryk Niewodniczański Institute of Nuclear Physics Polish Academy of Sciences 
contributes to the project, drawing on extensive experience in B-factory experimental environments since 1994. The Kraków group is involved in analyses of semileptonic and hadronic B decays with complex experimental signatures and large missing energy, which are challenging to study in hadronic collisions. Specific topics include:
(i) characteristics of semitauonic B decays, which test the interaction structure and lepton flavor universality. The first observation of exclusive $b \to c \tau \nu_{\tau}$ transitions in Belle was made by the Kraków group;
(ii) search for LFV in B meson decays using highly efficient, state-of-the-art reconstruction methods;
(iii) inclusive studies of B decays with the c s-bar quark system in the final state, including different charge combinations that are related to a specific production mechanism of the charm-strange meson. 

The Polish group contributes to the operation and software development of the Silicon Vertex Detector (SVD), a critical component for vertex reconstruction, flavor tagging, and reducing cross-feed backgrounds. These efforts include optimizing SVD performance during Run 2. Additionally, the group is actively participating in planning for the substantial detector upgrades required after Run 2 to ensure readiness for high-luminosity data taking in the next phases of the experiment. The Polish group’s efforts align with the global Belle II strategy, playing a vital role in advancing precision measurement capabilities and exploring new physics opportunities.

\subsection{Future CERN and other facilities}

LHCb plans to further expand its research program with future upgrades scheduled for implementation after 2030.
At CERN, the FCC-hh aims to develop a new particle accelerator capable of studying heavy-flavor phenomena at even higher energies than those currently accessible at the LHC.
Additionally, the EIC, currently under construction at Brookhaven National Laboratory in the USA, will play an important role in heavy-flavor physics by exploring hadronization processes—the formation of visible matter from quarks and gluons—through the reconstruction of hadrons and jets containing heavy flavors.

The NCBJ BESIII group is also conducting feasibility studies on hyperon decays for the proposed Super Tau-Charm Factory in China \cite{Achasov:2023gey}. The published studies by Polish scientists will serve as foundational references for these future investigations.

The future plans of the Kraków Belle II group include participation in the SuperKEKB upgrade to enable operation with polarized electron beam, as well as an ultra-high luminosity upgrade to 10$^{36}$ cm$^{-2}$s$^{-1}$ \cite{Forti:2022mti}.

One option under consideration for the future of particle physics in Europe is the ESS facility in Lund, Sweden \cite{Alekou:2022emd}. In particular, the NCBJ and UJ groups are involved in the preliminary planning of projects such as HIBEAM/NNBAR (neutron--antineutron oscillations) \cite{Abele:2022iml} or REDTOP (rare decays of $\eta/\eta'$ mesons) \cite{REDTOP:2022slw}. At present groups at NCBJ and UJ are interested in the projects.

\section{Strong Interactions}

Polish experimental groups involved in strong interaction physics take part of following experiments at the large research laboratories all over the word:
    ALICE, ATLAS and LHCb at the LHC at CERN,
    NA61/SHINE, COMPASS, AMBER and MUonE at the SPS at CERN,
    STAR at RHIC at BNL,
    HADES at FAIR at GSI.
Theory efforts are parallel to experimental interests. Both experimental and theoretical involvements are described below
followed by future prospects and goals.

\subsection{Current experiments and theory researches} 
\label{sec:si1}

In this section, we briefly describe all the experimental and theory research where the Polish community of strong interaction physics is involved. 

\textbf{ALICE} experiment (AGH, IFJ, NCBJ, PW, UŚ groups)
focuses on studying of quark-gluon plasma (QGP) in heavy-ion collisions and physics of dense systems of partons at LHC energies. Groups work on angular correlations of identified hadrons for studying hadronization process, studies of system size via Bose-Einstein and Fermi-Dirac correlations, photon emission during QGP formation, forward physics, 
UPC and multiplicity  correlations and fluctuations to study initial state of QGP~\cite{ALICE:2023hou,ALICE:2023csm,ALICE:2019xkq,ALICE:2019hno,ALICE:2023gcs,ALICE:2024yqj,ALICE:2024yvg}. They maintain ALICE FIT detector~\cite{ALICE:2023udb} (Project Leader from IFJ) and oversee the core analysis framework, the Event Display and various technical operations~\cite{ALICE:2019cox,ALICETPC:2020ann,ALICE:2022qhn}.

\textbf{ATLAS} experiment (AGH, IFJ, UJ groups) is focused on complementary studies
of quark-gluon plasma properties in heavy-ion collisions~\cite{ATLAS:2024qdu,ATLAS:2024jji}, forward physics~\cite{ATLAS:2022mgx,ATLAS:2023zfc,ATLAS:2020mve} (studies of elastic scattering, soft and hard diffractive and photon-induced processes, and responsible for the Roman Pot detectors: ATLAS-ALFA and ATLAS-AFP) in pp as well as parton density function (PDF) and saturation studies in UPC~\cite{ATLAS:2017fur,ATLAS:2019azn,ATLAS:2020hii} at LHC energies.

\textbf{LHCb} experiment (AGH, IFJ, UJ groups) concentrates analyses in the forward region of Pb-Pb, Pb-p, pp collisions at LHC energies. The greatest emphasis is placed on the dynamics of the system evolution (QGP) using Bose-Einstein and Fermi-Dirac correlations for various particle types, angular correlations of D-mesons and angular correlations of identified hadrons. Groups are interested in femtoscopic studies of particle correlations and diffractive central exclusive production (CEP)~\cite{LHCb:2017pnz,LHCb:2018rcm,LHCb:2023dcc}.

\textbf{NA61/SHINE} experiment (AGH, IFJ, NCBJ, PW, UW, UWr, UJ, UJK, UŚ groups) is a fixed target spectrometer at the SPS, operating in the collision energy range $5 < \sqrt{s}_{\mathrm{NN}} < 17$ GeV. It scans over various systems: p+p, p+C, $\pi$+C, K+C, p+Pb, Be+Be, C+C, Ar+Sc, Xe+La, and Pb+Pb reactions. Currently, the groups' focus is mostly on studies of open charm production in nucleus-nucleus collisions at SPS energies~\cite{NA61SHINE1a,NA61SHINE7}, isospin (flavour) symmetry violation studies in the charged over neutral K meson ratio~\cite{NA61SHINE:2023azp,NA61SHINE6}, and collision energy and system size dependence of hadron production which gives insight into the changeover from confined to deconfined matter~\cite{NA61SHINE2, NA61SHINE3, NA61SHINE4, NA61SHINE5,NA49}. 
The Polish community in NA61/SHINE is the largest group from Poland in all HEP experiments. Polish groups constitute about 50\% of the Collaboration and are responsible for key experimental activities.

\textbf{COMPASS} experiment (NCBJ, PW, UW groups) relies on high-energy muon and hadron beams from the M2 beam line at SPS. It works on both polarised and unpolarised proton and nuclear targets. The main goal is to investigate a 3-D structure of the nucleon through Transverse-Momentum Dependent (TMD) distributions and through Generalized Parton Distributions (GPD) and study hadron spectroscopy. The experiment has contributed with fundamental, first ever results on DIS, SIDIS, Drell-Yan and DVCS processes as well as hadron spectroscopy and polarisability.
The experiment has concluded its operations, data analysis is ongoing~\cite{aghasyan2017first}.

\textbf{AMBER} experiment (NCBJ, PW, UW groups) is in commissioning phase. The main goals are the measurements of cross sections for production of antiprotons in the proton-helium scattering within the energy range 50-280 GeV, which are crucial for dark matter searches, muon-proton elastic scattering at 100 GeV and very small $Q^2$ values to determine the proton charge radius and addressing an important missing piece in the proton charge radius puzzle, and parton distribution functions (PDFs) for pion and kaon mesons, explored through measurements of Drell-Yan and charmonium production using positively and negatively charged meson beams at 190 GeV.

\textbf{MUonE} experiment (IFJ group) is in commissioning phase~\cite{MUonE:2019qlm}. The main goal is to measure nonperturbative hadronic contribution to the anomalous muon magnetic moment.
The MUonE group from IFJ PAN is fully responsible for the central MUonE software based on FairRoot, including algorithms for all the crucial stages of experimental data analysis, i.e. event reconstruction, detector simulation, alignment and offline selection~\cite{MUonE:2021dqn,Kucharczyk:2019amx,Zdybal:2024yzu}.

\textbf{STAR} experiment (AGH, PW groups) was designed to discover and study the QGP properies. Currently it investigates the QCD phase diagram via Au-Au collisions and concentrate on studies of crossover transition between hadronic and quark matter, the first-order phase transition, and the hypothetical critical point. Moreover, the analysis of Beam Energy Scan (BES) data is ongoing to search for turn-off QGP signatures, signals of the first-order phase transition, QCD critical point, and signals of chiral symmetry restoration. It also operates with polarized proton beams to explore new territories related to the physics of spin~\cite{STAR:2005auv}.

\textbf{HADES} experiment (AGH, IFJ, PW, UJ, UŚ, UW groups) is a fixed-target experiment. It operates with ions (Au, Ag, C), proton or secondary pions beams of a few GeV provided by the SIS18 synchrotron. It probes QCD phase diagram in Au+Au, Au+C, C+C collisions in the energy range $0.8 < \sqrt{s_{NN}} < 2$~GeV~\cite{HADES:2009auv, HADES:2019auv, HADES:2021auv}. 

\textbf{Theory} (AGH, IFJ, NCBJ, PK, UJ, UJK, UMA, UŚ, UWr groups):
Polish QCD theory and phenomenology researches are focused on questions ranging from the behavior of QCD in high-density domains to the effects of higher-order corrections and the extraction of parton distribution functions of protons and nuclei, also using lattice QCD~\cite{Duwentaster:2022kpv,nCTEQ:2023cpo,Kovarik:2015cma,Alekhin:2014irh,xFitterDevelopersTeam:2018hym,xFitter:2022zjb,Lin:2017snn,Alexandrou:2018pbm,Alexandrou:2020zbe,Cichy:2019ebf}. The activity work covers low-$x$ QCD at high parton densities (IFJ, UJ, AGH, NCBJ)~\cite{vanHameren:2022mtk,vanHameren:2023oiq,Ganguli:2023joy,Caucal:2023fsf,Kotko:2023ugv}, quark-gluon plasma studies (IFJ, UJ, AGH)~\cite{Bozek:2022cjj,Samanta:2023qem,Bozek:2023dwp,Samanta:2023rbn,Parida:2023qju,Barej:2022ccb,Pei:2024wsy}, 3D imaging of hadron structure through exclusive and semi-inclusive processes (NCBJ), and entanglement in high-energy collisions (IFJ, UJ)~\cite{Hentschinski:2024gaa,Liu:2022bru}, with strong ties to experiments such as ALICE, and new facilities EIC, LHeC, and FCC-hh.  
Theoretical groups (IFJ, UJ, UW) are also developing Monte Carlo generators~\cite{Bellm:2015jjp,Bewick:2023tfi}, which are central to the research program on strong interactions at the LHC.

\subsection{Goals and perspectives}
\label{sec3}
Some experiments plan to continue data taking (e.g. HADES), some have already finished (e.g. COMPASS), and a few of them are planning to be upgraded (e.g. ALICE, NA61/SHINE). There are experiments already under construction (CBM at FAIR), planned ones in the advanced stage (e.g. ALICE 3), and some facilities (e.g. EIC, FCC-hh) to cope with new theoretical ideas or discover unexpected phenomena.

\textbf{Upgrades of ongoing experiments:} ALICE experiment plans to install the Forward Calorimeter~\cite{ALICE:2024jtt} to concentrate on the low-$x$ physics.
A key objective is to uncover gluon saturation, precisely determine PDFs and GPDs, 
determine PDFs using lattice QCD methods and explore phenomena like jet quenching in forward rapidity domain.

NA61/SHINE plans the installation of a new Liquid Hydrogen Target, a new Multi-gap Resistive Plate Chamber (MRPC) ToF detector, and a water target. The first motivation here is the measurement of (anti-)deuteron and multi-strange baryon production in high-statistics p+p interactions~\cite{NA61SHINE9}. The second motivation is a detailed study of particle production in light-ion collisions, aimed at uncovering the characteristics of the transition from confined to deconfined matter and studying the violation of isospin symmetry in charge-symmetric nuclear systems~\cite{NA61SHINE:2023azp,NA61SHINE10, NA61SHINE8}. Subsequently, the experiment plans to install a fast Large Acceptance Silicon Tracker (LAST) aimed at measurements of charm and anti-charm hadron correlations in Pb+Pb collisions to investigate the space correlation (locality) at their origin~\cite{Gazdzicki}. A TDR is in preparation.

\textbf{New detectors and facilities:}
There is a significant involvement in high rate CBM detector, already under construction at FAIR (AGH, IFJ, PW, UJ, UŚ, UW), which primary mission will be to understand strongly interacting matter in the region of  high baryochemical potential relvant for the structure of neutron stars, the dynamics of neutron star mergers, supernova explosions, the role of hyperons and hypernuclei in neutron stars, the nature of the phase transition from hadronic matter to quark-gluon matter, and to find experimental evidence for the restoration of chiral symmetry~\cite{CBM:2017auv, CBM:2020auv}.
The full silicon ALICE 3 detector~\cite{Adamova:2019vkf} (IFJ, AGH, PW, NCBJ express the forward detector contribution) at HL-LHC is proposed to study QGP in lower $p_{\mathrm{T}}$ range. LoI is accepted~\cite{ALICE:2022wwr}. 
There is a significant involvement into the EIC phenomenology~\cite{Adkins:2022jfp} (AGH, IFJ, NCBJ, PK, PW, UJ, UR, UW) and its already approved detector ePIC (AGH, IFJ - far forward and backward detectors) which will perform fundamental measurements of the nucleon and nucleus structure (including the polarised structure functions), and concentrate on the nucleon spin puzzle and non-linear QCD phenomena.
There is also an interest in LHeC (AGH) machine to study nucleon structure, nonlinear QCD or scan QCD phase diagram as well as in FCC-hh (AGH, IFJ, PWr, UJ, US) to study forward physics or study quark matter properties.
Theory long term plans address development of precision tools for higher-order calculations both in collinear and color glass condensate (CGC) factorization scheme as well as  precise determination of PDF using lattice QCD.

\section{Neutrino and Astroparticle Physics}

Polish neutrino physicists are involved in three groups of experiments studying different phenomena related to neutrinos: oscillations, with accelerator and atmospheric neutrinos, neutrinoless double beta decay and high-energy astrophysical neutrinos.

\subsection{Long baseline experiments}

The largest group is participating in neutrino experiments located in Japan: Super-Kamiokande (SK) \cite{Super-Kamiokande:2002weg} and T2K \cite{T2K:2011qtm}, which are currently taking data, and Hyper-Kamiokande (HK) \cite{Hyper-Kamiokande:2018ofw}, being under construction and expected to start in 2027. At present, the group consists of about 50 FTEs from nine institutes (NCBJ, UW, PW, CAMK, IFJ, UJ, AGH, UŚ, UWr).
There is a close collaboration between experimentalists, theorists and engineers. 

For almost 20 years, the Polish T2K group has been involved in the operation and research performed at the near detector ND280, starting with the participation in the ND280 construction and commissioning \cite{Aoki:2013swe,T2K:2022atj}. Several years ago, we took over the responsibility for the operation and calibration of the Fine Grained Detectors (FGDs), one of the sub-detector systems of the ND280, as well as FGD data quality assessment. Other ND280 tasks included developing track reconstruction algorithms, simulating external backgrounds, producing Monte Carlo simulations, and estimating several detector systematic errors.

Theorists’ efforts have focused on neutrino interaction models \cite{Kabirnezhad:2017jmf,Ankowski:2014yfa,Graczyk:2007bc}. The key advancements were made through the development of the NuWro neutrino interaction generator \cite{Golan:2012rfa} and significant contributions to the NEUT generator. At present, the attention is directed towards the implementation of the 2p2h interaction model in NuWro, which will also be adopted in NEUT. 

Polish physicists participate in the neutrino cross-section measurements performed at ND280, with a particular interest in channels with meson (pion or kaon) production \cite{Kowalik:2022hjt,Zarnecki:2022,grzegorz_zarnecki_2020_4254028}, which help to verify and improve neutrino interaction models. We plan to continue our work with the upgraded ND280  \cite{T2K:2019bbb}with better angular acceptance and lower momentum threshold for particle reconstruction. The detection of low-energy protons will be of particular interest for validating the cascade model used in NuWro. 

Polish group took part in the construction of two new High-Angle TPCs. The contribution refers in particular to the design and fabrication of the mechanical components, such as end plates and support structures for the TPC readout modules. We participated in the construction and testing of TPC prototypes, testing of a new type of Resistive Micromegas readout modules, in the assembly and testing of the TPCs at CERN, and in their installation in Japan. Some members of the group took part in the preparation of the SuperFGD module for installation, beam tests of the SuperFGD detector \cite{Blondel:2020hml} and played an important role in coordinating the work on the upgrade of the ND280, both in terms of mechanical work and safety aspects.

Polish group is involved in the flagship oscillation analyses as well \cite{T2K:2013ppw,T2K:2017hed,T2K:2019bcf}. The contribution includes the preparation of new ND280 event samples to better constrain the systematic uncertainties; new Far Detector event samples \cite{Prabhu:2024,SanjeevPrabhu:2023iem}; the development of the Bayesian fitter based on Markov chains \cite{Skwarczynski:2023,Skwarczynski:2022yxe}; performing fake data studies; validations and data fits. That activity included crucial contributions to T2K stand-alone analysis and T2K-NOvA joint analysis \cite{Nosek:2024ihz}.

Polish group recently increased engagement in the SK experiment, aiming to gain experience in operation, calibration and analysis performed in a Water Cherenkov detector, having in mind the future project, to which Poland contributes a lot: HK. In particular, Polish physicists were involved in T2K-SK joint analysis \cite{T2K:2024wfn} and recent studies of atmospheric neutrino oscillations \cite{Super-Kamiokande:2023ahc,Super-Kamiokande:2021the}, including tau appearance \cite{Mandal:2023rcy}.

Some of the Polish group members are also highly involved in the work for the WCTE (Water Cherenkov Test Experiment) project, which aims at operating a 40-ton test experiment in CERN. The list of Polish contributions to WCTE includes the construction, testing and characterization of the multi-PMT photon detectors, building the DAQ software, work on the front-end electronics and participation in the detector installation, calibration and operations. There are also plans to participate in the analysis of the data collected by the WCTE experiment and gain experience to contribute to the precise measurements at the planned Intermediate Water Cherenkov Detector (IWCD) for HK.

Polish institutes contribute directly to the preparation of the HK detector. Apart from the aforementioned work on multi-PMTs, a group of engineers works on the design of data concentration modules and low-voltage power supply modules for the underwater electronics of 50-cm PMTs, which will then be fabricated in Poland and installed on-site. Another group is responsible for the design, production and installation of an electron linear accelerator (together with the beam steering system and the deployment system), which will be used for calibration purposes. Poland also contributes to the purchase of PMTs and combined power and fiber optic cables. Polish physicists and engineers will work on the construction and commissioning of the detector systems on-site and provide the maintenance of the components built in Poland in the future.

The involvement also includes the work on neutrino interaction models and their implementation in NuWro, simulation of beam-related and cosmic background for IWCD and participation in large-scale production of Monte Carlo samples. The development of analysis methods and software is also planned, as well as participation in oscillation and cross-section studies after the start of the data taking.

There is also some interest and individual contributions to the neutrino short and long baseline experiments (SBN, DUNE) in the US, both based on LAr detectors. 

\subsection{NA61/SHINE measurements}

Measurements of key importance for the reduction of uncertainties on accelerator neutrino flux below 5\% are performed by the NA61/SHINE fixed-target spectrometer at the CERN SPS \cite{NA61SHINE:2015bad,NA61SHINE:2017fne}. These include studies of total production and differential ($p$, $\theta$) cross-sections of $\pi^\pm$, $K^\pm$, $p$, $\bar{p}$, $\Lambda$, $\bar{\Lambda}$ in collisions of different hadrons with thin targets \cite{NA61SHINE:2018hif,NA61SHINE:2019aip,NA61SHINE:2019nzr,NA61SHINE:2022uxp,NA61SHINE:2023bqo} as well as replicas of the T2K and NuMI targets \cite{NA61SHINE:2018rhe,NA61SHINE:2020iqu}. Currently, the main focus of the group is on measurements of hadron cross-sections from a prototype LBNF/DUNE target. Plans for Run 4 include the construction of a multipurpose low-energy beamline, aimed at reducing flux uncertainties for atmospheric neutrinos and bringing further improvement in accelerator neutrino studies \cite{NA61SHINE:SPSC-P-330-ADD-12,NA61SHINE_SPSC_M_793,NA61-lowenergy}.

Similarly, unique measurements with large phase-space coverage from NA61/SHINE relate to understanding air showers induced by ultrahigh-energy cosmic rays and tuning dedicated models. These include ($p$, $p_T$) spectra of identified particles and the production of V$^0$ particles and meson resonances in $\pi^-$+C interactions at 158--350 GeV/$c$ beam momentum~\cite{NA61SHINE:2017vqs,NA61SHINE:2022tiz}. Further measurements of high importance for Galactic cosmic-ray physics include nuclear fragmentation~\cite{NA61SHINE:2024rzv} to interpret recent high-precision spectra of secondary and primary cosmic-ray fluxes, and studies of nuclear coalescence in p+p collisions to calibrate the (anti)deuteron background in searches for DM annihilation signals (see~\cite{Maurin:2025gsz} for a recent review).\\
The overall involvement of the Polish NA61/SHINE community in these measurements accounts for about 20 people.

\subsection{Other neutrino experiments}

A group from the Jagiellonian University participates in the LEGEND \cite{LEGEND:2017cdu} and Borexino \cite{Borexino:2008gab} experiments. The LEGEND experiment has been designed to search for neutrinoless double beta ($0\nu\beta\beta$) decay in $^{76}$Ge by using HPGe detectors produced from enriched material. The Borexino detector was used to register solar neutrinos in real time.

Since the LEGEND collaboration formation in 2016 the main responsibilities of the UJ group were related to background reduction with appropriate screening techniques to select radio-pure materials (including design and construction of dedicated ultra-high sensitivity screening stations) as well as to development of the LAr purification techniques (improvement of the scintillation properties) and construction of appropriate equipment (purification systems for LAr) and to development of pulse shape discrimination methods (identification and rejection of background events). The first stage of LEGEND, LEGEND-200 (L-200, 200 kg of $^{76}$Ge) has presently the world leading sensitivity to the $^{76}$Ge $0\nu\beta\beta$ half-life ($10^{27}$ yr) and discovery. The experiment is ongoing at the Laboratori Nazionali del Gran Sasso (LNGS) in Italy. Simultaneously, LEGEND-1000 (L-1000, 1000 kg of $^{76}$Ge) is under preparation, which should reach the $0\nu\beta\beta$ decay half-life of $10^{28}$ yr. The construction of L-1000 at LNGS should start in 2026 and be completed in 2030.

Borexino Polish group was deeply involved in the background reduction in this experiment during its construction phase, and presently it focuses mostly on the data analysis and development of software methods for identiﬁcation and rejection of the background events. 

Polish groups are also involved in research with very high-energy neutrinos, oriented mostly on searches for high-energy neutrino sources in the sky and measurement of their flux’ properties. 

The team contributing to KM3NET \cite{KM3NeT-LoI-2.0} is responsible for software development and event reconstruction \cite{KM3NeT:2020tvi,gSeaGen_Alfonso_VLVNT2021,gSeaGen_Jutta_VLVnT2021,gSeaGen_repo,gSeaGen_Zenodo}. The work includes developing a simulation of underwater acoustic signals, to be applied to acoustic detector position calibration. The implementation of novel ML-based neutrino reconstruction via their acoustic signals is carried out and will allow for cosmic ray and neutrino studies at UHE energies.

The Polish group in the P-ONE experiment \cite{Malecki:2024tvt} from the INP PAS (IFJ PAN) in Kraków 
contributs to design and construction of a laser-based detector calibration system, design of analysis towards the early measurements (i.e. muon flux and its properties) and detector simulations (fast cascade generator, fast photon propagation algorithm, simulations of muon events).

Giant Radio Array for Neutrino Detection (GRAND): the experiment aims at building 20 arrays of 10,000 radio antennas each placed on 10,000 km$^2$ (altogether 200,000 km$^2$) in 2030s to detect Ultra-High-Energy neutrinos and cosmic rays \cite{GRAND:2024atu}. Currently GRAND operates 2 prototypes, one in the Gobi desert in China and one in Argentina, encompassing several dozen antennas. The Polish group, formed at the University of Warsaw, focuses on software and analysis, mainly offline event identification and Xmax reconstruction \cite{Piotrowski:2023xbc}. 

\subsection{Astroparticle physics}

One of the relevant topics in modern astrophysics is to understand the origin of cosmic-ray particles with energies above tens of EeV, and extremely low fluxes below one particle per km$^2$ per year, known as Ultra High-Energy Cosmic Rays (UHECRs). Dedicated to this task is the Pierre Auger Observatory (PAO), located in Argentina \cite{PierreAuger:2015eyc}. It detects particles from large atmospheric showers, which are cascades of secondary particles triggered by incoming UHECRs. Included in the APPEC Roadmap, PAO is a hybrid detector that combines a 3,000 km² network of water Cherenkov detectors with a network of telescopes observing the fluorescence emission from the showers. Polish scientists 
are focused on analyzing the optical images of large atmospheric showers and conducting research to identify photons and neutrinos among the cosmic ray particles \cite{Homola:2003ru,GORA2007402}. Currently, the PAO is undergoing upgrades in a~project known as Auger Prime, with active participation from Polish engineers, including the assembly and testing of surface scintillation detectors \cite{PierreAuger:2023yab}.

Polish scientists are also participating 
in the Cosmic Ray Extremely Distributed Observatory (CREDO) \cite{CREDO:2020pzy}, which was initiated at IFJ. CREDO employs a grassroots approach to scientific research, engaging citizen scientists. The main goal of CREDO is to search for signatures of cosmic ray ensembles, which consist of correlated large atmospheric showers and/or individual particles of secondary cosmic radiation, with a global network of detectors ranging from smartphones to professional infrastructure capable of recording cosmic rays as signal or background. Other forward-looking projects involving Polish scientists include JEM-EUSO \cite{2023EPJWC.28306007P} and POEMMA \cite{2021JCAP...06..007P}, which are planned as international space missions with super-wide-field telescopes designed to detect from space the fluorescence UV emission of atmospheric showers initiated by UHECRs above tens of EeV, as well as high-energy neutrinos above tens and hundreds of PeV. Currently, a~small orbital telescope, Mini-EUSO \cite{2025APh...16503057B}, is operating on the ISS, and a stratospheric balloon flight, POEMMA-Balloon with Radio (PBR), is in preparation \cite{2024NIMPA106969819B}. Polish team members
focus on data analysis, software development, and the preparation of high-voltage power supplies.

The Cherenkov Telescope Array Observatory (CTAO) project is dedicated to building a next-generation gamma-ray observatory operating within a wide range of very high energies (VHE), from approximately 20 GeV to 300 TeV \cite{2013APh....43....3A,2019APh...111...35A}. Since its inclusion in the ESFRI roadmap in 2008, the ASTRONET Roadmap 2022-2035, and the Polish Roadmap since 2020, CTAO aims to construct networks of Cherenkov telescopes of various sizes in both the Northern and Southern Hemispheres. These telescopes will detect atmospheric showers initiated by VHE gamma rays through their Cherenkov (optical) radiation, with an order-of-magnitude greater sensitivity than that of currently operational observatories \cite{2022Galax..10...21S,2022EPJST.231....3B}. At the beginning of 2025, CTAO became an ERIC, with construction scheduled for completion in 2032. Polish scientists 
from 13 institutions have contributed significantly to the development and scientific work of previous-generation Cherenkov telescopes --- H.E.S.S. \cite{2006A&A...457..899A} and MAGIC  \cite{2016APh....72...76A}--- for about two decades \cite[e.g.,][]{2022NatAs...6..689A,2024Sci...383..402H}. Their roles in CTAO include producing mirrors and mechanical structures for medium-sized telescopes, developing the Software Array Trigger (SWAT) for the Array Control and Data Acquisition (ACADA) system \cite{2024SPIE13101E..1DO}, developing low-level calibration procedures for large-sized telescopes \cite{2020SPIE11447E..0HN}, and participating in the development of software for the Science User Support System (SUSS) and the Science Operations Support System (SOSS). One of the flagship projects during the design phase of the CTAO was the construction of the Single-mirror Small-size Telescope (SST-1M) by Polish scientists in collaboration with Czech and Swiss teams \cite{2025JCAP...02..047A}. In this project, Polish units designed the mechanical structure of the telescopes including the drive, control, and positioning systems for the telescopes, mirror alignment components, and innovative electronics for digital triggering and data acquisition from cameras based on semiconductor photomultipliers \cite{2017NIMPA.845..350A,2017EPJC...77...47H}.
Polish scientists are also actively involved in constructing the Large-Sized Telescope prototype (LST-1), which is located at the CTAO-North site on the Canary Island of La Palma and is already operational \cite{2023ApJ...956...80A}.

Currently, the only detectors in the world conducting systematic observations of gravitational waves are those developed under the American LIGO project \cite{2015CQGra..32g4001L}, the European Virgo project \cite{2015CQGra..32b4001A}, and the Japanese KAGRA detector \cite{2013PhRvD..88d3007A}. These projects are coordinated through an agreement within the LIGO-Virgo-KAGRA Consortium (LVK), which is intended to eventually be transformed into a unified research infrastructure, the International Gravitational Wave Network (IGWN). Polish scientific involvement in the LVK consortium includes 
scientific staff, engineers, and doctoral students from ten Polish institutions.  
The contribution to the Virgo infrastructure 
included the design and construction of a network of seismic sensors around the Virgo detector, developing software packages for analyzing gravitational wave signals,  expanding a network of magnetometers to measure extremely low-frequency electromagnetic disturbances in LVK detectors \cite{PhysRevD.107.022004}, and establishing a Polish Tier-2 data analysis node within the IGWN Computational Infrastructure.

\section{Instrumentation}

\subsection{Contributions to the experiments on the LHC, HL-LHC, and other facilities}
\label{sec:inst_lhc}

Poland has made significant contributions to the LHC experiments, with various research groups involved in the design, construction, operation, and maintenance of critical instrumentation for both current experiments and future upgrades planned for the HL-LHC (High Luminosity LHC) phase. Below is a brief summary of the our contributions to each of the major experiments at the LHC:

Polish groups contribute significantly to several critical components of the CMS detector. Notably, they are responsible for the design and construction of the RPC readout systems \cite{Banzuzi:478904}, components of the Level-1 muon sub-trigger based on the comparison of RPC hits, and the OMTF (Overlap Muon Track Finder) trigger \cite{Bluj:2016cex}. The RPC control and data transmission system was also developed. This system is designed for the configuration and monitoring of RPC detectors, as well as for transmitting RPC data. Furthermore, there is ongoing work on the design of a muon trigger for the LHC Phase-II upgrade \cite{CERN-LHCC-2017-012}. There is also involvement in the electronic design of the front-end data processing system for the CMS High Granularity Calorimeter project \cite{CMS:2022lry} and front-end electronics for the MIP timing detector \cite{CMS:2667167}.
%
One should also note activ participation of the Polish teams in development of the forward physics PPS subsystem \cite{CMS:2021ncv}: detector calibration procedures, offline and data analysis software. In the phase-II upgrade of the PPS subsystem the contributions are related to the frontend electronics and detector development.

Polish contributions to the ATLAS detector include the power supply system for the Semiconductor Tracker \cite{Malecki:530702}, the Gas System, and the Detector Control System for the Transition Radiation Tracker \cite{10.1117/12.2000242}, as well as the ALFA detector \cite{AbdelKhalek:2016tiv}. Polish teams are also involved in the Phase-II Upgrade of the ATLAS detector \cite{ATLASCollaboration:2012ilu}. The Polish teams are contributing to the upgrades of two subsystems: the Inner Tracker (ITk) \cite{Gonella:2023uag} and the Trigger and Data Acquisition (TDAQ) system \cite{2137107}. In the ITk projects, the key areas of involvement include the design of front-end ASICs for the readout of silicon strip detectors, the high-voltage and low-voltage power supply systems for these detectors and the front-end electronics, and the cooling system for the ITk. Specifically, the team is responsible for developing custom radiation-tolerant DC-DC converters for the low-voltage power supply system, as well as highly specialized components for the cooling system. In the TDAQ system, the Polish teams contribute to the development of hardware and firmware for the jFEX and L1Topo modules for the Phase-I upgrade, as well as the jFEX and L0Topo modules for the Phase-II upgrade.

For the ALICE experiment, Polish groups are responsible for the simulation and calibration of the Time Projection Chamber (TPC) \cite{Lippmann:2014lay}, the high-granularity Photon Spectrometer (PHOS) \cite{ALICE:1999ozr}, the Fast Interaction Trigger (FIT) \cite{Trzaska:2017reu}, FoCal forward calorimeter \cite{CERN-LHCC-2024-004}, and the development of the diffractive detector AD (ALICE Diffractive) \cite{VillatoroTello:2017ccr}. 

In the LHCb experiment, Polish groups have made significant contributions to the Outer Tracker \cite{580712}, Vertex Locator (strip sensors) \cite{Collins:2013pu}, Upstream Tracker \cite{LHCb:2014uqj}, and High-Level Trigger subsystems \cite{Witek:2012gb,LHCb:2003aa}. They are also actively involved in the LHCb Upgrade I \cite{LHCb:2023hlw} and consolidation projects, contributing to the Vertex Locator (pixels) Upstream Tracker \cite{Kopciewicz:2022gfo}, RICH detectors, Magnet stations. 
For the Upstream Tracker Polish groups took the whole responsibility for the design and production of the 128-channel readout ASIC called SALT \cite{Beteta:2021gpd}, which is already installed and used in the data taking. 
In the next Upgrade II phase \cite{LHCb:2021glh}, Polish groups are involved in the Scintillating Fibres (SciFi) and in the Magnet Stations. 

In addition to the contributions to individual LHC experiments, Polish group have made significant contribution to the lpGBT ASIC project \cite{Moreira:2025iiq} – a fast data concentrator for all detectors as part of the upgrades of all LHC experiments and many others around the world.

Polish contributions at CERN are not only related to the LHC.
Polish groups have a significant impact on the following experiments at CERN: COMPASS/AMBER \cite{Adams:2018pwt} (SciFi, FEE for ECAL0, upgrade of CEDAR), NA61/SHINE \cite{NA49-future:2006qne} (reconstruction, calibration, DRS system, DAQ), AEgIS \cite{AEGIS:2012eto} (Control System, plasma monitoring, electrostatic beamlines, negative/positive ion sources), GBAR \cite{Perez:2015zya} (LINAC for positron production),  
ISOLDE (extensive and long-term exploitation), CLICdp \cite{CLICdp:2018vnx} (FCAL and pixel detectors R\&D) and MuonE (detector layout and technology).

Numerous contributions to instrumentation were also made outside CERN. For Belle II experiment at SuperKEKB \cite{Belle-II:2018jsg} they included SVD as well as pixel detectors R\&D, protection and LV systems. For the upgrade of T2K ND280 detector \cite{T2KND280TPC:2010nnd}:  TPC mechanical support and testing of MicroMegas, assembly and testing of TPCs at CERN, and installation at J-PARC, beam test, calibration and installation of Super Fine Grained Detectors (FGDs), reconstruction algorithms, and the construction of two new High-Angle TPC detectors.
Super-K \cite{Super-Kamiokande:2002weg} and Hyper-K \cite{Hyper-Kamiokande:2018ofw} contributions include design, and construction of multiPMTs for the far and intermediate detector as well as design of data concentration modules and low-voltage power supply modules for the underwater electronics of 50\,cm PMTs. 
There are plans to work on the construction and commissioning of the detector systems in Japan and provide the maintenance of the components built in Poland. Works also include the design, production and installation of an electron linear accelerator (with the beam steering system and the deployment system) for calibration purposes. Polish group is currently highly involved in WCTE (Water Cherenkov Test Experiment) project \cite{Barbi:2712416} including the construction, testing and characterization of the multi-PMT photon detectors, DAQ software, Front-end electronics. Participation in the detector installation, calibration and operations.

For the LUXE experiment at Eu-XFEL (DESY, Hamburg) \cite{Abramowicz:2021zja,LUXE:2023crk} Polish groups are involved in the development and construction of the positron calorimeter, ECALp, as well as calorimeter simulation and optimization studies. 
%
Polish FAIR in-kind contributions include: (i) cryogenics systems for SIS100 \cite{Eisel_2015} and Super Fragment Separator \cite{Roux:2024zwh} (ii) diagnostic systems for testing SIS100 dipole magnets  (iii) HADES Forward Tracker \cite{Rathod:2020jtz} with dedicated front-end and digital electronics (iv) sensors, front-end electronics and read-out firmware for the silicon tracking detector in CBM experiment \cite{RodriguezRodriguez:2025lor} (v) Forward Tracker (FT) based on straw tubes \cite{Smyrski:2018snv} with front-end and digital electronics \cite{Firlej:2023paa} in PANDA.
For accelerator and detectors at EIC, plans are to build a calorimetry system for bremsstrahlung photons measurements and to monitor synchrotron radiation, including front-end electronics development.
The Polish hardware contribution to the P-ONE Experiment \cite{P-ONE:2020ljt} includes laser calibration system development and its adaptation to P-ONE requirements, analysis design,   simulations for muon observations with the first line, development and validation of fast photon propagation simulation, analysis of STRAW data, and GEANT4 detector simulations.
For GERDA \cite{GERDA:2012qwd} and LEGEND \cite{LEGEND:2017cdu} Polish team are involved in various activities related to background reduction of the detector (crucial parameter of the detector), studies of LAr properties, optimization of the LAr scintillation yield.
Contributions to JEM-EUSO/Poemma project \cite{JEM-EUSOColaboration:2014pci,POEMMA:2020ykm} include High Voltage Power supply, hardware integration, and calibration, software, data analysis, and simulations, and for Pierre Auger Observatory (PAO) \cite{PierreAuger:2015eyc}: 
assembly and testing of surface scintillation detectors for the Auger Prime upgrade; 

For the Cherenkov Telescope Array Observatory (CTAO) \cite{CTAO:2024wvb} Polish units contribute to 
producing mirrors for Medium-Sized Telescopes,  developing the Software Array Trigger for the Array Control and Data Acquisition system, developing low-level calibration procedures for Large-Sized Telescopes and analysis software (joint LST1—MAGIC observations), development of software for the Science User Support System and the Science Operations Support System.
They are also responsible for construction of the Single-mirror Small-size Telescope (SST-1M) mechanical structure of the telescopes including the drive, control, and positioning systems for the telescopes, as well as mirror alignment components, and innovative electronics for digital triggering and data acquisition from cameras based on semiconductor photomultiplier. 

At the European Spallation Source (ESS) Polish teams are members of the HIBEAM/NNBAR experiment \cite{Burgman:2024rkm}. The contributions in instrumentation consist of two main subjects: the first is the design and construction of neutron spin polarizer and neutron spin polarimeter for the first phase of HIBEAM experiment where the possible coupling of neutron spin to the hypothetical magnetic-like field created by the axion-like dark matter halo in our galaxy will be examined. Both polarizer and polarimeter will utilize the spin-dependent neutron transmission through the polarized $^{3}$He. The second is the design and construction of large dimension, high-performance cosmic veto detectors (CVD) for HIBEAM, still in the first phase of the project, and for NNBAR experiments. They will consist of a multi-layer structure of plastic scintillator slabs read out by arrays of SiPM light sensors attached to wave-length-shifting optical fibers. CVD as a whole must provide close to 100 \% detection efficiency for all foreseen species in broad energy range and position resolution for trajectory reconstruction permitting 100 \% discrimination against not-target vertices. 
In both cases the design processes are accompanied with intensive simulations of neutron and neutron spin transport in magnetic field, cosmic ray interaction with detection setup and ambient as well as beam related background. Multi-step, iterative prototyping are indispensable as well.

\subsection{Involvement in DRD Collaborations}

Following the recomendations of the ECFA Detector R\&D Roadmap~\cite{Detector:2784893},
Detector Research and Development Committee (DRDC) was formed in June 2023 by the CERN Director of Research to receive and evaluate proposals for new detector R\&D experiments. After evaluation and approval of applications prepared by the interested parties, eight DRD collaborations were established begining of 2024: DRD1 on Gaseous Detectors \cite{Colaleo:2885937}, DRD2 on Liquid Detectors \cite{Monroe:2886644}, DRD3 on Solid-State Detectors \cite{Cartiglia:2901958}, DRD4 on Photon Detectors and PID \cite{CERN-DRDC-2024-001}, DRD5 on Quantum and Emerging Technologies \cite{Doser:2901426}, DRD6 on Calorimetry \cite{CERN-DRDC-2024-004}, DRD7 on Electronics and On-Detector Processing \cite{Baudot:2901965}, DRD8 on Integration \cite{Schmidt:2922888}.
Based on the previous experience with R\&D programmes created and maintained at CERN (like RD50 and RD51) as well as other R\&D projects, Polish teams got involved in all but two DRD collaborations (no Polish groups are yet involved in DRD4 and DRD8). 

DRD1 is focused on the development of gaseous detectors, including Micro-Pattern Gas Detectors (MPGD) and Gaseous Detectors (GD). Modern GDs are suitable for a variety of applications in fundamental research domains and beyond. Their importance in particle physics experiments continues to be crucial, as evidenced by their incorporation into all major LHC experiments and numerous other experiments conducted at CERN and worldwide, which primarily use extended GD systems. Moreover, novel concepts are being developed within these experiments. DRD1 will focus on the development of new GD technologies, including the development of new MPGDs, such as Micromegas, GEM, and THGEM, and the development of new GDs, such as RPCs, Micromegas, and GEMs. Environmentally friendly development is a key element of the R\&D process, especially in the area of gas mixture studies, where not only detector performance is crucial but also the negative impact on the environment is minimised. Groups from Poland (AGH, CAMK, IFPiLM, UJ) contributing to DRD1 have broad experience in the field of gas studies for GDs.

DRD2 is focused on liquid detectors that primarily employ target media of water, liquid scintillator, or cryogenic noble liquids as Ar, Xe, or He. The detectors are used in addressing fundamental open questions in science today across neutrino physics, dark matter searches and astro-particle experiments.
Polish groups (CAMK, UJ) are involved in characterizing noble liquid target properties and in radiopurity and
background mitigation techniques.

DRD3 is focused on the development of Solid-State Detectors (SSD). SSDs are based on semiconductors, and in particular silicon detectors (planar pixels, planar strips, and 3D pixels), which are used in almost all particle physics experiments and beyond. Revolutionary improvements in SSD performance are needed to match the requirements of future experiments, regardless of which ones are selected to be built. A common future requirement for the front-end electronics is to perform very complex tasks, such as those required for 4D tracking or by the transfer off-chip of very large data volumes. 3D stacking is, therefore, a key technological development that needs to be included in future high-performing trackers. Groups from Poland (AGH, IFPiLM) have been actively participating in detector technology
and readout electronics development for a long time. Their existing experience strengthens the DRD3
research.

DRD5 is focused on advancing measurement concepts within the domain of quantum sensing. The main difference with respect of other DRD collaborations is the low initial technology readiness levels of the involved technologies. The collaboration will approach the incorporation of quantum sensing techniques into accelerator-based experiments. 
As an example of activities withing the DRD-5 collaboration, we can mention the R\&D towards developing superconducting sensors with 1\,ps timing resolution, with the capacity of reading 100\,Gcps to keep the integrated backgrounds negligible, highly granular with pixels of 1\,µm, and more environmentally friendly.
Polish groups (NCBJ, PW, UMK) contribute to all work packages and bear leading roles in several of them.

DRD6 is focused on developing radiation-hard calorimeters with enhanced electromagnetic energy and timing resolution, highly granular calorimeters with a multi-dimensional readout for optimised use of particle flow methods, and calorimeters for extreme radiation, rate and pile-up environments.
Polish groups (AGH, UW) contribute to the development of Sandwich Calorimeters (i.e. alternating absorption and sensitive layers) and to the development of electronics for calorimeters.

DRD7 is devoted to developing future electronic systems and technologies for particle physics detectors. Polish group (AGH) contributes to the development of ultra-low power high-performance TDC and ADC blocks.

DRD4 and DRD8 do not have Polish participation yet. However, given the initial stage of the collaborations, it is highly likely that groups from Poland will join in the future as tasks related to photon detectors and integration of detectors have been implemented by Polish teams in the past in different experiments

\section{Theory}

Theoretical high-energy physics research in Poland focuses on the following topics: (i) preparing Monte
Carlo Event Generators for direct use by the experimental collaborations, (ii) investigating extensions of the Standard Model (SM) and testing them against experimental data, (iii) performing precision calculations within the SM, with special attention to processes where signals of new physics could occur, (iv) studying QCD dynamics in bound states and in high-energy multipartonic processes, (v) studying strongly interacting matter at high temperatures and densities, (vi) analyzing theoretical aspects of quantum field theories and string theory.

The main research centres are located in Kraków, Warszawa and Wrocław, sizeable ones in Katowice, Kielce and Łódź, while smaller groups are active in Białystok, Poznań, Rzeszów, Szczecin and Zielona Góra. 
%
Short descriptions of the topics (i)-(vi) are given below.

(i) Monte Carlo Event Generators (MCEGs) that simulate particle collisions, or ``events'', to a level that allows direct comparison with experimental data, are indispensable for the planning and analysis of past, current, and future experiments in particle physics. MCEGs bridge the gap between relatively abstract theoretical concepts or calculations and the often complex experimental reality. They provide inputs to detailed software models of detectors, are key components in the evaluation of systematic uncertainties on measurements and frequently play an essential role in the interpretation of those measurements. As such, MCEGs support the ongoing success of an experimental program through continuous improvement, addressing ever-increasing demands for their precision, coverage, flexibility and usability. The development, maintenance, validation and calibration of these central assets is in large part driven by researchers across European universities, constituting a relatively small but critically important ecosystem of authors of MCEGs and supporting tools for their deployment, validation and calibration. Polish research groups have a long and well-known tradition in preparing (or participating in preparation) of widely used MC simulation programs, both for hadronic and e$^+$e$^-$ colliders. Examples of such codes are CARLOMAT \cite{Jegerlehner:2017kke}, EKHARA \cite{Czyz:2018jpp}, EpIC \cite{Aschenauer:2022aeb},
GF-CAIN \cite{Bieron:2021ojp}, HDECAY \cite{Djouadi:2018xqq}, KaTie \cite{vanHameren:2016kkz}, MadGraph5\!\_\!aMC@NLO \cite{Flore:2025ync,Wettersten:2025hrb}, MINCAS \cite{Kutak:2018dim}, PHOKARA \cite{Campanario:2019mjh}, PHOTOS \cite{Davidson:2010ew,Golonka:2005pn}, TAUOLA \cite{Jadach:1990mz,Jadach:1993hs}, WHIZARD \cite{Kilian:2007gr,Moretti:2001zz,Reuter:2024dvz}, WINHAC \cite{Placzek:2003zg,Placzek:2013moa}, general-purpose Monte Carlo event generators Herwig \cite{Bellm:2015jjp,Bellm:2019zci} and Sherpa \cite{Sherpa:2024mfk} or the family of codes including KORALZ \cite{Jadach:1999tr}, KrkMC \cite{Sarmah:2024hdk,Jadach:2015zsq}, BHLUMI \cite{Jadach:1991by,Jadach:1996is}, BHWIDE \cite{Jadach:1995nk}, KKMCee \cite{Jadach:1999vf,Jadach:2022mbe}, KoralW \cite{Jadach:1998gi,Jadach:2001mp}, YFSWW \cite{Jadach:1995sp,Jadach:2001mp}, YFSZZ \cite{Jadach:1996hh}, based on the YFS method of calculating exactly the electroweak processes via soft photon resummation. In the case of neutrino interactions, the NuWro MC generator~\cite{Prasad:2024gnv} is one of the major simulation tools used by experimental groups studying neutrino oscillations. It is a complete generator with detector interface covering neutrino energy
range from $\sim$100~MeV to $\sim$100~GeV. The ongoing physics exploitation at the (HL)LHC and other current experiments, and the preparation for future facilities such as Higgs Factory, Gamma Factory or Muon Collider, mandates the continuation of the strategic and vibrant research programme of this community at the interface of experiment, theory, and computation. Such an inter-sectional character of the MCEG developers' activity poses serious challenges in terms of career paths and funding opportunities, and it is imperative that the particle physics community addresses such issues in order to ensure a sustainable development of MCEGs for existing and future facilities.

(ii) BSM studies that aim at investigating complete models require including constraints from accelerator experiments, astrophysical observations, dark matter detection, neutrino mixing, and low-energy precision experiments. Additional requirements that are often imposed include the hierarchy problem, resolving flavour physics puzzles, successful baryogenesis, proper description of perturbations during the cosmological inflation, etc. Usually, only a subset of the above constraints is taken into account. For instance, noticeable effort is devoted to extended Higgs sector models that provide acceptable dark matter freeze-out and correct electroweak baryogenesis at the same time~\cite{Grzadkowski:2018nbc}. Their common property is the existence
of new particles that are within reach either for the upgraded LHC or for the planned e$^+$e$^-$ colliders (ILC, CLIC, FCC-ee). Another path is supersymmetric models with $\mathcal O$(1~TeV) higgsinos as dark matter candidates~\cite{Kowalska:2018toh}, offering partial explanations to observed tensions in anomalous magnetic moments of the leptons and
several B-meson decays, which is of relevance for Belle II and LHCb. Neutrino physics studies cover both the Dirac and Majorana mass scenarios, collider probes and tests of neutrino mass-generation models, often requiring extra discrete or continuous symmetries, and allowing for new sources of CP violation~\cite{Kim:2022xjg,Ganguly:2022qxj,Dziewit:2023dzg}. A model-independent approach to BSM physics is carried out within the Standard Model Effective Field Theory (SMEFT) where Wilson coefficients of non-redundant higher-dimension operators (e.g., the
so-called Warsaw basis~\cite{Grzadkowski:2010es}) parameterise effects of heavy particles on physics around the electroweak scale and below. Polish groups are also involved in the development of the CheckMATE framework~\cite{Dercks:2016npn} for reinterpreting physics searches at the LHC and are involved in data and analysis preservation efforts.

(iii) Multi-loop perturbative calculations within the SM are being performed for the purpose of precision electroweak measurements at the future e$^+$e$^-$ colliders~\cite{Jadach:2019wol,Dubovyk:2019szj}, precision two- and three-loop QCD and resummations calculations at LHC, vector boson scattering and fusion, as well as for measurements of rare B-meson decays~\cite{Bobeth:2013uxa,Misiak:2015xwa} at LHCb and Belle II. They often involve thousands of Feynman diagrams, for the
generation of which codes like FeynRules~\cite{Dedes:2023zws} are developed. The diagrams are evaluated using various techniques, including algebraic reduction, differential equations and Mellin-Barnes representations~\cite{Dubovyk:2022frj}. In the case of differential spectra with cuts, when phase-space integration is an issue, development of devoted Monte Carlo codes is necessary.

(iv) Phenomenological investigations of QCD dynamics in Poland encompass a broad range of topics. These include structure functions and parton density functions, higher-twist effects, forward physics at the LHC, parton saturation, sum rules, double parton distributions, and diffraction, jet substructure and quark-gluon
jets as well as multiparticle production and determination of parton densities in the proton and nuclei within the nCTEQ and the xFitter projects~\cite{Duwentaster:2022kpv,nCTEQ:2023cpo,Kovarik:2015cma,Alekhin:2014irh,xFitterDevelopersTeam:2018hym,xFitter:2022zjb,Lin:2017snn,Alexandrou:2018pbm,Alexandrou:2020zbe,Cichy:2019ebf}.
Development of non-perturbative QCD methods and models such as hadronization and color reconstruction models. Applications of machine learning techniques in QCD, for example, to the hadronization process or the calculation of entanglement entropy using neural networks.
Correlations and geometrical scaling are also studied, along with the relationship between entanglement entropy and parton density functions. Effective models of QCD confinement, grounded in chiral symmetry, are being developed and applied to research on the QCD phase diagram, particle spectroscopy in light and heavy-light sectors, and the dynamics of light meson interactions in the non-perturbative region up to approximately 2~GeV. These efforts also include calculations of parton distributions, both generalized and transverse-momentum-dependent, and the development of beyond-leading-power factorization methods, such as kT factorization at next-to-leading order or Color Glass Condensate at next-to-eikonal precision.
Studies extend to form factors and transition form factors at low-energy scales. Lattice and light-front approaches are employed to explore QCD bound-state dynamics, including non-perturbative properties of the nucleon and pion (e.g., form factors and generalized parton distributions), ground-state formation in non-perturbative relativistic QCD, planar expansion, and dimensional reductions. Exclusive photoproduction of quarkonia in ultraperipheral collisions at the LHC, with a focus on the impact of nuclear gluon distributions, is under investigation, as is the theoretical description of hadron electroproduction at the proposed electron-ion collider (EIC) and the LHeC.

(v) As far as the strongly interacting matter at high temperatures and densities is concerned, activities of the Polish theory community include application of relativistic hydrodynamics for phenomenological description of heavy-ion collisions at RHIC and LHC as well as lower energy experiments~\cite{Florkowski:2017olj}. They combine the language of quantum field theory, relativistic kinetic theory, and holography as well as studies of particle correlations in heavy ion collisions. Other recent issues of interest are simulations of QCD cascades in quark-gluon plasma in heavy-ion collisions using Markov Chain Monte Carlo methods, and heavy quarks propagating in turbulent quark-gluon plasma, in which case the Fokker-Planck transport equation is being used~\cite{Mrowczynski:2017kso}. Temperature oscillations and sound waves in hadronic matter are investigated on the basis of LHC heavy-ion collision data on transverse momenta distributions~\cite{Wilk:2017coo}. Additionally, the theoretical investigation of exotic mesons, such as glueballs and hybrids, focuses on understanding their formation and decay processes.

(vi) Examples of Polish expertise in the hep-th domain include analyses of Donaldson-Thomas invariants, torus knots and lattice paths at the classical and quantum level~\cite{Kucharski:2017ogk,Panfil:2018sis}, studies of kappa-deformations of the Poincaré algebra~\cite{Borowiec:2015wua,Lukierski:2016vah}, conformal defects in supergravity~\cite{Janik:2015oja}, or appearance of infinite-dimensional symmetry groups in $N=8$ supergravity~\cite{Meissner:2018gtx}. A large part of Polish theoretical high-energy physics research is carried out in collaboration with foreign institutions, often within well-established international teams. Some of these
teams are led by physicists from Poland.

\section{Accelerator components and technologies}

Polish scientific institutions and industry have successfully participated in developing, constructing, and commissioning of Big Science infrastructure for several decades. Their expertise in accelerator and detector technologies has led to significant contributions to major international projects, such as e.g. CERN’s Large Hadron Collider (LHC, HiLumi LHC), European XFEL, European Spallation Source (ESS), Facility for Antiproton and Ion Research (FAIR), PIP II (FERMILAB) and numerous other cutting-edge scientific endeavours. The accumulated experience in superconducting accelerator technologies positions Poland as one of key players in upcoming global accelerator projects. The competences developed and maintained in Poland cover most of advanced components and technologies of superconducting Big Science accelerators. The contributions from Poland include a wide range of specialized activities, demonstrating Poland’s expertise in accelerator science and engineering. The main areas of involvement can be categorized as follows:

\textbf{Cryogenic Distribution Systems:} The Polish cryogenic distribution systems are in operation or under construction in XFEL, ESS, FAIR, CERN and FERMILAB \cite{Duda_2017,Arnold_2024,Polinski_2022,Eisel_2015,Venturi_2020,Banaszkiewicz_2024_1,Martinez_2024}. The systems comprise up to several tens of valve-boxes, span over hundreds of meters and are characterized by state of art thermo-mechanical parameters. The systems have been fully designed in Poland and predominately manufactured in Polish industry.

\textbf{Cryomodules comprising Superconducting Magnets and RF Systems:} The Polish scientific institutions in collaboration with industry have integrated and cryomodules with superconducting (SC) magnets, radio frequency (RF) systems and diagnostics components. High competences regarding LLRF systems in designing hardware and all layers software modules for stabilizing accelerating fields with high precision. Advanced control techniques and fast frequency tuning, ensuring optimal performance in beam acceleration in large-scale facilities such as FLASH, X-FEL, and ESS.

\textbf{Superconducting Magnet and Circuit Testing:} Polish teams have played a crucial role in the verification of electrical circuits and interconnections in superconducting magnets, particularly at CERN’s LHC and the SIS100 cryo-magnetic system at FAIR.

\textbf{Precise System Synchronization and RF Instrumentation:} Development of electronic systems, instrumentation, and software for particle accelerators, focusing on RF front-end circuits, beam instrumentation, and Low-Level RF (LLRF) control.

\textbf{Instrumentation and Quality Assurance:} Extensive involvement of Polish institutions in verification of superconducting magnet circuits, assembly of detector cooling systems, and prototyping of measuring equipment. Involvement in establishing and handling local databases, programming measuring equipment, and developing RF power compressors in projects like XFEL and PIP-II. Design, installation, and commissioning of infrastructure for beam instrumentation at ESS, including rack and patch panel design, documentation, test procedures, and EMC solutions. Risk analysis of the accelerator systems and novel instrumentation developments. 

\textbf{Beam Dynamics:} Polish researchers have conducted beam simulations, studied beam lifetimes, and developed precision luminosity measurement tools for projects such as LHeC, FCC, T2K and PIP-II. 

\textbf{Assembly and Installation of Detector and Accelerator Components:} Contributions include assembling inner detector cooling systems at CERN experiments, installing power supplies, and fabricating plasma diagnostic equipment at W7-X.

\textbf{Development of RF Synchronization Systems:} Polish scientists contributed to the development and installation of highly reliable 1.3 GHz Main Oscillator systems for FLASH and European XFEL, achieving sub-10 fs stability.

\textbf{RF Testing and Installation:} Polish experts have been responsible for the testing and installation of RF distribution and phase stability systems, power converters, and cryomodules at ESS, XFEL, and FAIR. Development of a "Cavity Simulator" for ESS, the first device capable of real-time simulation of an entire RF station, enables control system conditioning without requiring access to accelerating modules. 

\textbf{Advanced Data, Electronics and Control Systems:} Involvement in establishing and handling local databases, programming measuring equipment, and developing RF power compressors in projects like XFEL and PIP-II. Design and deployment of multiple devices for LLRF control, such as MTCA.4 RTM backplanes, multichannel downconverters, LO generation modules, and RF vector modulators, contributing to the operation of machines such as FLASH, XFEL, PETRA, ESS, and CERN projects. High competences in xTCA-based systems, digital and analog electronics for control applications, protection systems (as RF Protection Interlock for PIP-II), firmware development, and accelerator control system software. These technologies have been successfully deployed in LLRF systems for projects such as X-FEL, ESS, and PolFEL, improving system reliability and operational precision.

\textbf{Industrial and Medical Accelerator Applications:} Development of high-intensity linear accelerators for medical applications and participation in multi-billion euro global supply chains.

\textbf{Engineering and Prototyping of Tools:} Polish engineers have designed and manufactured critical tools for accelerator assembly and maintenance, including custom tools for W7-X and SPIRAL2.

Research in Poland is advancing along multiple directions in accelerator science. One area of focus is beam dynamics and collider instrumentation, with Polish scientists contributing to the conceptual design of the proposed LHeC electron-hadron collider at CERN. These efforts involve state-of-the-art beam dynamics simulations and research into beam lifetimes and beam-beam interactions for future circular electron-positron colliders, including studies on beamstrahlung and bremsstrahlung interplay. Additionally, novel instrumentation for colliders is being developed, particularly for precise luminosity and beam-position monitoring systems.

The \textbf{Gamma Factory}~\cite{GF2015,GF2022} is a scientific project developed as part of the Physics Beyond Colliders (PBC) research programme. Its purpose is to produce high-intensity gamma beams (outperforming the currently available gamma-ray sources by at least 7 orders of magnitude) in laser beam interactions with highly ionised heavy atoms accelerated in SPS and LHC accelerators. 
These types of gamma rays can then be used to produce intense (surpassing the currently available sources by several orders of magnitude) beams of polarised electrons, polarised positrons, polarised muons \cite{Apyan:2022ysh}, neutrinos, neutrons, and radioactive ions \cite{Nichita:2021iwa}. It is a unique gamma-ray factory -- possible to be realised only at CERN -- which opens up wide prospects for research in many fields of science, such as particle physics \cite{Placzek:2024bkv}, nuclear physics, atomic physics, accelerator physics \cite{Krasny:2020wgx}and astrophysics, and also creates great opportunities for many application solutions, e.g. in the field of nuclear energy \cite{Baolong:2024ata} and nuclear medicine. 
An experiment is currently being prepared to demonstrate the feasibility (Proof-of-Principle) of the Gamma Factory project using the SPS accelerator \cite{GF-PoP-LoI:2019} -- planned for the years 2026--2027.
The Polish contribution to the Gamma Factory project includes precise calculations of energy levels and lifetimes in excited states for highly ionised heavy atoms, theoretical predictions of the interaction of laser light with highly ionised atom beams, especially in the form of a Monte Carlo generator ({\sf GF-CAIN}) for simulation of relevant physical processes, as well as expertise in the field of laser physics \cite{Bieron:2021ojp}.

Polish researchers are also engaged in accelerator manufacturing and their technological development. The expertise includes the design and fabrication of electron linear accelerators for scientific, medical, and industrial applications, as well as the production of dipole and quadrupole magnets, warm accelerating structures, and cooling systems. Monte Carlo simulations for shielding, target optimization, and radiation protection studies are an integral part of ongoing research. Furthermore, projects such as calibration systems for Hyper-Kamiokande, FLASH radiotherapy, RF power compressors and optimization of superfluid helium cryogenic systems show Poland’s growing capabilities in advanced accelerator technologies.

The extensive expertise accumulated by Polish institutions and industry over decades has positioned Poland to play a critical role in future accelerator projects. Key areas for future engagement include: 
(i) participation in the Future Circular Collider (FCC) and muon collider initiatives,
(ii) advancements in Energy Recovery Linacs (ERLs) and plasma-based accelerators,
(iii) expansion into medical accelerator technologies and high-intensity linear accelerators.

A particularly important project for the future of Polish accelerator research is the \textbf{Polish Free Electron Laser (PolFEL)}. PolFEL is designed as a state-of-the-art facility \cite{Banaszkiewicz_2024_2,Nietubyc_2021} that will provide high-brilliance, ultra-fast pulses of THz to UV light for a wide range of scientific applications. This project is crucial not only for expanding Poland’s expertise in superconducting accelerator technology but also for maintaining a highly skilled workforce and strengthening collaboration with international partners. The development of PolFEL, despite its relatively small scale, integrates all key accelerator technologies, including cryogenics, state of art cryomodules with superconducting cavities, RF power systems, and advanced beam diagnostics, making it a national priority for future scientific and industrial applications. Its successful construction and operation will position Poland among the leading contributors to free-electron laser research worldwide.

\section{Computing}

National Research and Educational Network PIONIER, operated by the Poznan Supercomputing and Networking Center (PSNC), plays a critical role in providing the high speed required to support HPC computing resources in Poland. The PIONIER network's infrastructure offers robust, high-speed, and scalable connectivity that is essential for the computing ecosystem in Poland, including data-intensive tasks, distributed computing, and seamless collaboration across national and European partners. At the national level, PIONIER’s full-mesh architecture connects all major Polish HPC data centers and distributed data storage sites with direct, high-speed links. This setup delivers low-latency communication and efficient data transfer across Poland, real-time data analysis, and resource-sharing among institutions. With bandwidth capacities of nx400 Gbps, and even 800 Gbps through dedicated lambdas, the PIONIER network and core services ensure robust support for the processing of massive datasets. 
On an international scale, PIONIER connects seamlessly to major European research and education hubs, such as Hamburg, Amsterdam, and Geneva, through high-speed links that integrate it into the GEANT network, the backbone of European research collaboration. Additionally, PIONIER facilitates efficient IP traffic exchange through key domestic and international nodes, such as PLIX, AMS-IX, DE-CIX, and LINX, ensuring smooth global data transfer and reliable communication at the minimum 100 Gbps level. 
Three computing centers, the Academic Computer Centre ACK Cyfronet AGH in Kraków, the National Centre for Nuclear Research - Świerk Computing Center (NCBJ), and the Poznan Supercomputing and Networking Center (PSNC), provide resources for LHC computing (as the Polish Tier2 of WLCG). In addition, the NCBJ Swierk Computer Center provides computing resources supporting LHCb experiment data as Tier 1.
Polish high-energy physics (HEP) research groups have access to supercomputing resources through the PLGrid infrastructure, which unites the six largest supercomputing centers in the country. In addition to ACK Cyfronet AGH, PSNC, and NCBJ, the federation includes the CI TASK, ICM UW, and WSNC computing centers. Research groups can also utilize smaller computing clusters for local and grid purposes, such as those at IFJ PAN in Kraków and the Tier 3 center at NCBJ in Warsaw, which operates using WLCG software. The resources from these centers are shared by several other HEP experiments, including Auger, Belle, COMPASS, CTA, Lofar, HESS, ICARUS, NA61, SuperB, T2K, and P-ONE. 

In April 2021, the “Memorandum of Understanding for Collaboration in the Deployment and Exploitation of the Worldwide LHC Computing Grid (WLCG).” AGH's Academic Computer Center Cyfronet (Cyfronet), the National Center for Nuclear Research (NCBJ), and the Institute of Bioorganic Chemistry of the Polish Academy of Sciences' Poznan Supercomputing and Networking Center have concluded an agreement to share computer resources for data processing for Poland's obligations in the WLCG.                             	 
The signatories to the Agreement are three Polish data centers that will make specific commitments to the WLCG initiative. These include commitments to make hardware and personnel resources available and to maintain them for a declared period. The agreement concerns sharing resources for experiments at the Large Hadron Collider (LHC) at the European Organization for Nuclear Research (CERN): ALICE, ATLAS, CMS, and LHCb. 
The provision of resources for individual experiments is carried out and coordinated by specific Signatories to this Agreement, in coordination with the coordinators of each experiment, and so:
ACK Cyfronet AGH provides and coordinates resources for the ATLAS collaboration,
NCBJ  for the LHCb collaboration and
PSNC for the ALICE collaboration.

Since the beginning of data collection through experiments at the Large Hadron Collider (LHC) and beyond, the demand for computing power and storage capacity has steadily increased. Over the years of LHC operations, the average annual growth in computing resources at Tier 2 cluster levels across all experiments has been approximately 20\%.
A significant operational break for the LHC accelerator is planned for 2026, during which modernization and upgrades to the machine and its detectors will be carried out. Once operations resume in 2030, the LHC will operate with higher luminosity, resulting in a substantial increase in the number of recorded events. During this phase, data collection is expected to increase by approximately tenfold compared to previous operational phases~\cite{CERN-LHCC-2022-005,Software:2815292}. Consequently, maintaining an annual 20\% growth in computing resources at Polish computing centers is critical to fulfilling Poland's commitments to the LHC experiments.
Sustainable and consistent development of IT infrastructure is essential to accommodate the expected surge in computing power and data volumes to ensure active participation in future experiments.

The attractive proposition is to improve computing centers by investing in GPGPU accelerators. Following the huge worldwide increase in usage of AI methods in recent years, there have been several approaches to utilize these methods in HEP as well. Especially quick Monte Carlo simulations can benefit greatly from them. As Monte Carlo simulations constitute a large portion of CPU resource usage, the AI revolution may have a significant influence on the HEP computing landscape. Despite current, very high, unit prices of GPGPU accelerators, they are usually cheaper and consume less energy per FLOPS than traditional CPU architectures. Nevertheless, there are also some risks associated with AI-driven GPGPU product development. As the AI revolution is focused mostly on Large Language Models (LLMs), there is big demand for higher throughputs in low-precision computations. Therefore emerging products tend to have much higher numbers of low-precision (FP32, INT8) computing units that cannot be used as parts of double-precision (FP64) computing units as well. Therefore HEP applications, which usually require double precision, may not fully benefit from current market trends. 

Currently, Poland primarily supports Tier 1 and Tier 2 computing centers, which provide essential resources for LHC experiments.
However, the computing resources delivered for non-LHC experiments have almost been exceeded. Historically, many research institutions involved in LHC and other experiments also operated smaller Tier 3 clusters. These clusters played a crucial role in supplementing declared resources and supported local research groups during the final stages of their analyses before publishing results.
Unfortunately, most Tier 3 clusters have been disconnected from the Grid due to outdated equipment or insufficient maintenance funding. Reviving these Tier 3 clusters would provide a cost-effective way to enhance resource availability within the network and increase computing power for the local research groups This approach would enable Poland to meet its computing obligations more effectively with relatively modest financial investments. \\

In conclusion, the current allocation of computing resources for high-energy physics research in Poland is sufficient to meet its obligations for the time being. However, it is essential to secure continuous financial support and ensure an annual increase in resources of around 20\% in the long term. Without these measures, Poland may struggle to fulfill its commitments to both the LHC and non-LHC experiments. The accumulated resources following the LHC era should be adequate for future research initiatives like the FCC, new linear colliders,  EIC, and many others.

\clearpage

\printbibliography 

\end{document}